\PassOptionsToPackage{table}{xcolor}
\documentclass{article} % For LaTeX2e
\usepackage{iclr2027_conference,times}

\usepackage{amsmath,amsfonts,bm}

\def\eqref#1{equation~\ref{#1}}
\def\1{\bm{1}}

\DeclareMathAlphabet{\mathsfit}{\encodingdefault}{\sfdefault}{m}{sl}
\SetMathAlphabet{\mathsfit}{bold}{\encodingdefault}{\sfdefault}{bx}{n}

\usepackage{hyperref}
\usepackage{url}
\usepackage{booktabs}       % professional-quality tables
\usepackage{amsfonts}       % blackboard math symbols
\usepackage{nicefrac}       % compact symbols for 1/2, etc.
\usepackage{microtype}      % microtypography
\usepackage{xcolor}         % colors
\usepackage{amsmath}
\usepackage{graphicx}
\usepackage{multirow}
\usepackage{enumitem}
\usepackage{wrapfig}
\usepackage{multicol}
\usepackage{amsthm}
\usepackage{tcolorbox}
\usepackage{wrapfig}
\tcbuselibrary{breakable}
\usepackage{subcaption}
\usepackage{float}
\usepackage{graphicx} 
\usepackage{algorithm}
\usepackage{algpseudocode}
\usepackage{caption}
\usepackage{graphicx}
\usepackage{amsmath,amssymb}  
\usepackage{amsthm} 
\usepackage{array}
\usepackage{makecell}
\usepackage{mathtools}
\usepackage[table]{xcolor}
\usepackage{ulem}
\usepackage{textcomp}
\usepackage{tabularx}
\definecolor{headergray}{RGB}{232,232,232}
\definecolor{sectiongray}{RGB}{247,247,247}
\definecolor{hermesblue}{RGB}{226,239,246}
\definecolor{bestblue}{RGB}{205,229,238}
\definecolor{secondorange}{RGB}{246,224,198}
\definecolor{gainbg}{RGB}{226,240,218}
\definecolor{lossbg}{RGB}{247,224,224}

\newcommand{\best}[1]{\cellcolor{bestblue}\textbf{#1}}
\newcommand{\second}[1]{\cellcolor{secondorange}\textbf{#1}}
\newcommand{\gain}[1]{\cellcolor{gainbg}\textbf{#1}}
\newcommand{\loss}[1]{\cellcolor{lossbg}\textbf{#1}}

\setlist[itemize]{noitemsep,leftmargin=*,topsep=0pt}
\title{Harness Engineering for Software Engineering via Modular Executable Dev-Primitives}

\author{
\begin{tabular}{@{}l@{}}
Haibo Jin$^{1}$\quad
Xinjie Li$^{2}$ \quad
Peng Kuang$^{1}$ \quad
Haohan Wang$^{1}$\thanks{Corresponding Author: \texttt{haohanw@illinois.edu}}  \\[3pt]
{\small $^{1}$University of Illinois Urbana-Champaign, USA} \\
{\small $^{2}$The Pennsylvania State University, USA}
\end{tabular}
}
\iclrfinalcopy
\begin{document}

\maketitle

\fancyhead{}
\lhead{Preprint.}

\begin{abstract}
Large language models (LLMs) equipped with terminal access have demonstrated strong capabilities in automating software engineering tasks. However, existing agents remain brittle on long-horizon workflows, where they must repeatedly reconstruct program state scattered across source files, configurations, tests, dependencies, and runtime behavior, leading to increasingly long interaction histories, context explosion, and semantic drift. Large repositories further complicate the identification of task-relevant components. To address these challenges, we introduce \textbf{Dev-Primitives} (\emph{Development Primitives}), a modular and executable abstraction that transforms repository components from passive software artifacts into active participants in software engineering. Each Dev-Primitive pairs a repository artifact with a resident LLM, which gives the artifact an agent-native interface grounded in its own implementation and dependencies, enabling natural-language reasoning, inter-component communication, and localized self-modification. Building on Dev-Primitives, we propose \textbf{HERMES}, a \textbf{H}arness \textbf{E}ngineering framework for software enginee\textbf{R}ing via \textbf{M}odular \textbf{E}xecutable Dev-Primitive\textbf{S}, which instantiates these primitives at repository scale through a dependency-aware dynamic activation mechanism and a bug diagnosis mechanism that maps execution evidence back to the components that must be revised. Extensive experiments on four software engineering benchmarks demonstrate that HERMES outperforms matched baseline harnesses by 12.4\% on average. Moreover, when paired with strong activation and diagnosis models, HERMES, even with Qwen3-8B Dev-Primitives, remains within 4.5\% of the homogeneous GPT-5.6 Sol configuration across all four benchmarks, while reducing inference cost by 26.2\% on Terminal-Bench 4.0, highlighting the importance of harness design in software engineering agents.

\end{abstract}

\section{Introduction}
Large language models (LLMs) equipped with terminal access have demonstrated remarkable capability in automating real-world software engineering~\citep{yang2024swe, wang2025openhands}. Existing approaches broadly follow two paradigms. \emph{Pipeline-based methods} prescribe the control flow in advance, decomposing software repair into stages such as fault localization, patch generation, and validation~\citep{xia2024agentless, zhang2024autocoderover}. \emph{Agent-based methods} instead place an LLM in an interactive execution environment, where it can inspect repositories, execute commands, modify code, and react to runtime feedback through a general action loop~\citep{wang2024executable, wang2025openhands}. Recent benchmarks have meanwhile expanded from isolated issue resolution to tasks that require sustained interaction with repositories, terminals, tests, and runtime systems~\citep{merrill2026terminal}.

Both paradigms remain brittle once such interactions extend over long horizons. Append-only histories and passive compression lead to context explosion, semantic drift, and degraded reasoning~\citep{liu2026context, wang2026improving}. In software repositories, however, the difficulty is not only the length of the history: the relevant program state is scattered across interdependent source files, configurations, and tests, so a change to one file often requires matching changes in the code that calls it, its configuration, and its tests. Each edit also changes the repository itself, so the agent must repeatedly re-inspect affected files and work out what each change requires elsewhere, a burden that no amount of history compression removes. Recent evaluations expose this clearly: on whole-repository migration, only 5.4\% of 520 agent runs complete all evaluation stages~\citep{hong2026swe}, and similar difficulties appear in long-horizon software evolution and repository generation~\citep{le2025swe, ding2025nl2repo}. This raises a natural question: \textbf{\emph{instead of requiring an agent to repeatedly reconstruct distributed program state, can individual software components reason about their own responsibilities and communicate relevant information directly to one another?}}

We answer this question with \textbf{Dev-Primitives} (\emph{Development Primitives}), modular and executable abstractions that transform repository components from passive code artifacts into active participants. Each Dev-Primitive wraps a repository artifact with a \emph{resident LLM}, and this is what makes the artifact agent-native: because the resident model reads the implementation it is attached to, the component can be addressed in natural language and can answer for its own implementation and dependencies. Through this interface it interprets requests, reasons over its own artifact, communicates requirements and constraints to other primitives, and modifies its implementation when necessary, so that component-specific information and reasoning stay with the artifact that owns them.

Dev-Primitives alone, however, do not determine which components a task should involve: repositories may contain thousands of components, and activating all of them would be wasteful. We therefore introduce \textbf{HERMES}, a \textbf{H}arness \textbf{E}ngineering framework for software enginee\textbf{R}ing via \textbf{M}odular \textbf{E}xecutable Dev-Primitive\textbf{S}, which adds two mechanisms. \emph{Dynamic activation} localizes the components likely to implement the reported behavior, follows their dependencies to the callers, configurations, and tests that may need to change with them, and instantiates only the Dev-Primitives in this candidate set, which modify their artifacts and exchange requirements before the repository is executed. Because a single round often leaves the issue unresolved, and execution reports only \emph{that} the repository still fails rather than \emph{which} component should be revised, a \emph{bug diagnosis} mechanism traces the failure back to the components it implicates and reactivates only those primitives, repeating until the repository passes or a revision budget is exhausted.

Extensive experiments on four software engineering benchmarks show that HERMES consistently improves over existing harnesses across issue resolution, whole-repository refactoring, terminal-based tasks, and DevOps workflows. Under the default GPT-5.6 Sol setting with medium reasoning effort, it improves performance by 12.4 percentage points on average. With strong activation and diagnosis models, HERMES using Qwen3-8B Dev-Primitives stays within 4.5 percentage points of the homogeneous GPT-5.6 Sol configuration while reducing inference cost by 26.2\% on Terminal-Bench 4.0. These results demonstrate the importance of harness design in software engineering agents. Our contributions are summarized as follows:

\begin{itemize}
\item We introduce \textbf{Dev-Primitives}, modular executable interfaces that turn repository components into active software engineering participants capable of localized reasoning, natural-language inter-component communication, and self-modification.
\item We propose \textbf{HERMES}, a \textbf{H}arness \textbf{E}ngineering framework for software enginee\textbf{R}ing via \textbf{M}odular \textbf{E}xecutable Dev-Primitive\textbf{S}, which instantiates Dev-Primitives at repository scale through a dependency-aware dynamic activation mechanism and a bug diagnosis mechanism that maps execution evidence back to the components that must be fixed.
\item We evaluate HERMES on four software engineering benchmarks spanning issue resolution, whole-repository refactoring, terminal-based tasks, and DevOps workflows. HERMES improves over matched baseline harnesses by 12.4 percentage points on average, while Qwen3-8B Dev-Primitives remain within 4.5 percentage points of the homogeneous GPT-5.6 Sol configuration and reduce Terminal-Bench 4.0 inference cost by 26.2\%.
\end{itemize}

\section{Related Work}
\textbf{Software Engineering Agents and Long-Horizon Evaluation.}
SWE-agent~\citep{yang2024swe}, OpenHands~\citep{wang2025openhands}, and CodeAct~\citep{wang2024executable} place an LLM in an executable environment where it can inspect repositories, edit code, run commands, and react to runtime feedback, while Agentless~\citep{xia2024agentless} and AutoCodeRover~\citep{zhang2024autocoderover} prescribe explicit localization and repair stages, and other systems distribute work across designer-specified roles or file-level developers under centralized coordination~\citep{chen2024coder,liu2024marscode,tao2024magis,wang2026codeteam}. Evaluation has meanwhile moved toward sustained terminal interaction and multi-stage software workflows~\citep{merrill2026terminal,tang2026devops}, and toward repository-scale consistency during software evolution, repository generation, and whole-repository migration~\citep{ding2025nl2repo,hong2026swe}, where context growth, semantic drift, and state preservation become bottlenecks for long-running agents~\citep{liu2026context,wang2026improving}.

\textbf{Reusable and Modular Agent Capabilities.}
One line of work learns software engineering behavior from repository-level trajectories~\citep{pan2024swegym,yang2026swe,ma2024lingma,ma2025p2t}. Another represents reusable capabilities explicitly through modular abstractions~\citep{jin2026agent,jin2026harness,qiu2025agentdistill,li2025codeskill}, encapsulating reasoning procedures, skills, or tool interfaces that an agent can invoke and compose across tasks.

\textbf{Key Differences.}
These systems differ from Dev-Primitives in the unit to which reasoning capability is attached. Prior agents assign work to designer-specified roles or centrally coordinated file-level tasks, so the reasoning units and the paths along which they exchange information are fixed in advance~\citep{tao2024magis,wang2026codeteam}; prior primitives encapsulate reusable, task-agnostic behaviors~\citep{jin2026agent,jin2026harness,li2025codeskill}; and context-management methods keep information inside a single trajectory~\citep{liu2026context,wang2026improving}. A Dev-Primitive is instead defined by the artifact it owns, so the set of primitives is induced by the repository and requirements flow along its dependency edges, and HERMES activates these primitives on demand and revises them from execution evidence. Appendix~\ref{appendix:extended-rw} gives an extended discussion.

\section{Methodology}\label{meth}
\subsection{Overview}
\begin{figure*}[htbp]
    \centering
    \vspace{-10pt}
    \includegraphics[width=1\linewidth]{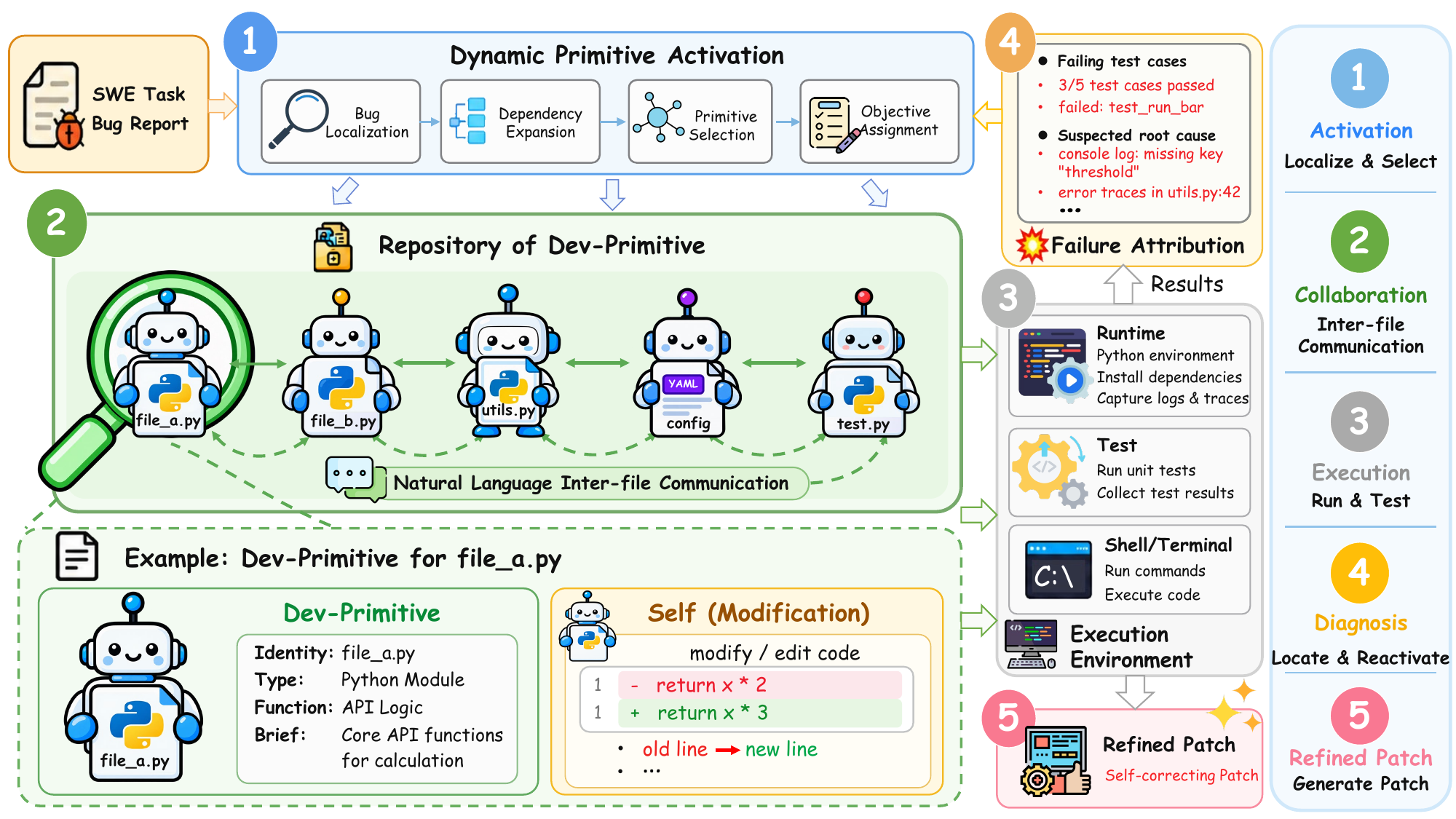}
    \vspace{-20pt}
     \caption{Overview of HERMES. Dynamic activation selects the repository components relevant to the task and instantiates Dev-Primitives, which modify their own artifacts and exchange requirements in natural language. The modified repository is executed, and bug diagnosis maps the observed failure back to the components that must be revised, reactivating only those for the next round.}
    \label{pipline}
    \vspace{-10pt}
\end{figure*}

HERMES is built on \textbf{Dev-Primitives}, modular and executable interfaces that transform repository components into active participants in software engineering. Since only a small subset of components is relevant to a given issue, HERMES instantiates Dev-Primitives on demand through a \emph{dynamic activation mechanism} that localizes likely components and expands along their dependencies. The activated primitives modify their own artifacts and are validated in an execution environment. Because execution reveals only that the repository still fails, HERMES additionally employs a \emph{bug diagnosis mechanism} that maps the observed failure back to components implicated by the evidence, so that each revision round reactivates only those primitives. An overview of HERMES is shown in Fig.~\ref{pipline}.

\subsection{Dev-Primitives}

A Dev-Primitive wraps an individual repository component, such as a source file, configuration file, or test file, with a \emph{resident LLM}. Each primitive has direct access to the artifact it represents and is responsible for reasoning about and modifying that artifact during software engineering. Unlike conventional agents that repeatedly retrieve repository files as passive context, Dev-Primitives turn these components into active participants that can interpret natural-language requests, communicate implementation requirements with other components, and directly edit their own artifacts.

Formally, let a repository consist of $N$ software components
$\mathcal{R}=\{a_i\}_{i=1}^{N}$. Each component $a_i$ is associated with a Dev-Primitive $P_i$ instantiated with an LLM $\mathcal{M}$:
\begin{equation}
    (a_i', m_i) = P_i(a_i, x_i, \mathcal{C}_i) = \mathcal{M}([a_i; x_i; \mathcal{C}_i]),
    \label{eq:devprimitive}
\end{equation}
where $x_i$ denotes the task assigned to the primitive, $\mathcal{C}_i$ denotes information received from other Dev-Primitives, $a_i'$ is the optionally modified artifact, and $m_i$ is a natural-language message containing task-relevant information or requirements to be communicated to other components. When no modification is required, $a_i'=a_i$.

A Dev-Primitive therefore performs \emph{local modification}, reasoning over and editing the artifact it represents, and \emph{inter-primitive communication}, exchanging task-relevant information with other Dev-Primitives through natural language.

\textbf{Inter-primitive communication.} The message $m_i$ carries what the primitive discovered while reasoning over its own implementation and what it therefore requires of other components: interface changes, implementation requirements, dependency updates, configuration constraints, or testing requirements. Messages are directed rather than broadcast. A primitive sends only to the components its local reasoning implicates, which are in practice its callers, callees, configurations, and tests, so the communication topology follows the repository's dependency structure rather than a predefined organization. Messages produced in a round enter the communication context $\mathcal{C}_j$ of their recipients before that round terminates, so a primitive may revise its own modification after receiving a requirement from another component. Participation does not imply modification: a primitive may contribute information about its implementation while leaving its own artifact unchanged.

% Because each primitive can interpret the artifact it owns, a requirement discovered during modification is addressed directly to the component that must satisfy it, rather than passing through an intermediate representation of the repository assembled by a central agent. Whether this is preferable to exposing the same components to a single editing agent is an empirical question, which we test in Section~\ref{exp} and Appendix~\ref{appendix:primitive-analysis}; Appendix~\ref{appendix:case-study} analyzes a complete trajectory with the messages exchanged.

\subsection{Harness Engineering via Modular Executable Dev-Primitives}

Only a small subset of a repository is relevant to a given issue, so instantiating every Dev-Primitive would be wasteful. \textbf{HERMES}, a \textbf{H}arness \textbf{E}ngineering framework for software enginee\textbf{R}ing via \textbf{M}odular \textbf{E}xecutable Dev-Primitive\textbf{S}, therefore supplies two mechanisms that the abstraction itself does not provide (Fig.~\ref{pipline}); complete specifications of each mechanism are in Appendix~\ref{appendix:component-specs}.

\textbf{Dynamic Primitive Activation.} Given an issue $q$ and repository $\mathcal{R}$, HERMES produces a plan $\Pi = \{(i,x_i)\}_{i\in\mathcal{A}} = \operatorname{ACTIVATE}(q,\mathcal{R})$, where $\mathcal{A}$ indexes the selected components and $x_i$ is the local objective assigned to $P_i$. It localizes the components likely to implement the reported behavior and expands along their dependencies to callers, callees, configurations, and tests, inspecting repository structure and retrieving files as needed rather than loading the repository into one context. Only $\{P_i \mid i\in\mathcal{A}\}$ are instantiated, so cost follows the size of $\mathcal{A}$ rather than that of the repository.

\textbf{Dev-Primitive Collaboration.} The activated primitives then operate as a round. Each receives its local objective $x_i$ from the plan together with the messages $\mathcal{C}_i$ addressed to it by other activated primitives, reasons over its own artifact in light of both, and produces $(a_i', m_i) = P_i(a_i, x_i, \mathcal{C}_i)$: an optionally modified artifact and an outgoing message carrying the requirements that its modification imposes elsewhere. Messages circulate only among the primitives in $\mathcal{P}_{\mathcal{A}}$ and reach their recipients before the round terminates, so a requirement discovered while editing one component can still be satisfied by the components that depend on it within the same round. Cross-file changes are coordinated this way without any single context holding every modified component at once.

\textbf{Execution Environment.} Rather than relying on model-side reasoning about correctness, HERMES evaluates the modified repository $\mathcal{R}'$ in an executable environment, producing $o = \operatorname{EXECUTE}(\mathcal{R}')$: command outputs and exit codes, test outcomes, runtime errors, and available logs and stack traces. These signals expose inconsistencies that appear only once independently modified components are exercised together. Note that held-out evaluation tests are never used during solving (Appendix~\ref{appendix:implementation-details}).

\textbf{Bug Diagnosis.} A single round of modification often leaves the issue unresolved, and execution then reports only whether the repository still fails, not which component should be revised. Bug diagnosis supplies this attribution, returning $v = \operatorname{DIAGNOSE}(q,\Pi,\mathcal{R}',o) \in \{\textsc{pass},\textsc{fail}\}$ and, when $v=\textsc{fail}$, structured feedback $\phi = (e, c, u)$: the observed failure, the suspected root cause together with the components implicated by the evidence, and revision guidance. The component set named in $c$ is what keeps revision localized. HERMES revises the plan as $\Pi' = \operatorname{ACTIVATE}(q,\mathcal{R}',\Pi,\phi)$, which may retune objectives, drop components, or activate ones the new evidence implicates, and the loop continues until the repository state is accepted or a budget of $B$ rounds is exhausted.

\section{Experiments}\label{exp}
\subsection{Experimental Setup}\label{setup}

\textbf{Benchmarks.}
We evaluate HERMES on four benchmarks spanning complementary software engineering settings. SWE-bench Verified~\citep{jimenez2023swebench} covers real-world GitHub issue resolution with human-validated tasks; SWE Refactor Bench~\citep{hong2026swe} targets whole-repository migration and coordinated cross-file modification; Terminal-Bench 4.0~\citep{merrill2026terminal} measures performance on interactive terminal-based tasks; and DevOps-Gym~\citep{tang2026devops} spans build and configuration, monitoring, issue resolution, test generation, and end-to-end multi-stage workflows.

\textbf{Baselines and Implementation Details.}
We follow the official leaderboard settings and evaluation protocols of each benchmark, and run HERMES with the backbones used by the published configurations under each baseline's matched setting. Our evaluations use seven backbones: GPT-5.6 Sol, Terra, and Luna~\citep{openai2026gpt56}, Claude Sonnet 5~\citep{anthropic2026claudesonnet5}, Claude Opus 5~\citep{anthropic2026claudeopus5}, DeepSeek-V4~\citep{xu2026deepseek}, and Qwen3-8B~\citep{qwen3technicalreport}. Unless otherwise specified, we use medium reasoning effort where configurable and a revision budget of $B=3$. In every result table, each group compares a baseline harness with HERMES on the identical backbone and reasoning effort; only the harness differs, and $\Delta$ is the percentage-point gain over that row's baseline; HERMES is run once per backbone, Qwen3-8B is evaluated with HERMES only, and blue and orange mark the best and second-best value per column. Backbone assignments, execution isolation, primitive runtime, ablation settings, and baseline provenance are given in Appendix~\ref{app:models} and Appendix~\ref{app:impl-eval}.

\begin{table*}[ht]
\centering
\vspace{-10pt}
\caption{Performance and efficiency on SWE-bench Verified. Resolved rates are mean $\pm$ standard error over 500 instances.}
\vspace{-10pt}
\label{tab:swebench-verified}

\Large
\setlength{\tabcolsep}{4.5pt}
\renewcommand{\arraystretch}{1.0}
\setlength{\aboverulesep}{0pt}
\setlength{\belowrulesep}{0pt}
\setlength{\extrarowheight}{1pt}

% group row: baseline harness on the left, HERMES on the right
\newcommand{\harnessrow}[1]{%
  \rowcolor{sectiongray}
  \textit{Harness} &
  \multicolumn{3}{c}{\textbf{#1}} &
  \multicolumn{3}{c}{\cellcolor{hermesblue}\textbf{HERMES}} &
  \\}

\resizebox{\textwidth}{!}{%
\begin{tabular}{l ccc ccc c}
\toprule

% ---------- header row 1 ----------
\rowcolor{headergray}
&
\multicolumn{3}{c}{\textbf{Baseline Results}}
&
\multicolumn{3}{c}{\cellcolor{hermesblue}\textbf{HERMES Results}}
&
\\
\cmidrule(lr){2-4}\cmidrule(lr){5-7}

% ---------- header row 2 ----------
\rowcolor{headergray}
\multirow{-2}{*}{\textbf{Backbone}}
&
\rule{0pt}{2.8ex}\textbf{Resolved (\%) $\uparrow$}
&
\rule{0pt}{2.8ex}\textbf{Cost/Test (\$) $\downarrow$}
&
\rule{0pt}{2.8ex}\textbf{Latency (s) $\downarrow$}
&
\cellcolor{hermesblue}\rule{0pt}{2.8ex}\textbf{Resolved (\%) $\uparrow$}
&
\cellcolor{hermesblue}\rule{0pt}{2.8ex}\textbf{Cost/Test (\$) $\downarrow$}
&
\cellcolor{hermesblue}\rule{0pt}{2.8ex}\textbf{Latency (s) $\downarrow$}
&
\multirow{-2}{*}{\makecell{\textbf{$\Delta$ Res.}\\[-1pt]\textbf{(p.p.)} $\uparrow$}}
\\

% ================= block 1: mini-SWE-agent =================
\midrule
\harnessrow{mini-SWE-agent}

GPT-5.6 Sol   & \second{96.20 $\pm$ 0.86} & 1.15 & 182.37 & \best{97.00 $\pm$ 0.76}   & 1.80 & \second{205.34} & \gain{+0.80} \\
GPT-5.6 Terra & 95.40 $\pm$ 0.94          & 0.40 & 180.00 & \second{96.20 $\pm$ 0.86} & 0.98 & \best{196.71}   & \gain{+0.80} \\
GPT-5.6 Luna  & 93.00 $\pm$ 1.14 & \best{0.04} & 201.07 & 95.60 $\pm$ 0.92 & \best{0.10} & 218.46 & \gain{+2.60} \\

Claude Fable 5  & 95.00 $\pm$ 0.98        & 2.05 & 356.18 & 96.00 $\pm$ 0.88        & 4.50 & 381.27 & \gain{+1.00} \\
Claude Opus 5   & \best{97.00 $\pm$ 0.76} & 1.29 & 576.99 & \best{97.00 $\pm$ 0.76} & 2.25 & 472.18 & 0.00 \\
Claude Sonnet 5 & 79.60 $\pm$ 1.80        & 1.49 & 962.37 & 85.80 $\pm$ 1.56        & 2.10 & 824.63 & \gain{+6.20} \\

DeepSeek-V4  & 77.40 $\pm$ 1.87 & 0.44 & 634.61 & 82.80 $\pm$ 1.69 & 0.83 & 418.52 & \gain{+5.40} \\
GPT-5.5      & 82.60 $\pm$ 1.70 & 1.36 & 426.43 & 85.60 $\pm$ 1.57 & 2.45 & 398.74 & \gain{+3.00} \\
GPT-5.4 Mini & 73.00 $\pm$ 1.99 & 0.51 & 326.46 & 82.40 $\pm$ 1.70 & 0.83 & 347.82 & \gain{+9.40} \\
GPT-5 Mini   & 60.80 $\pm$ 2.19 & \second{0.05} & 187.13 & 80.20 $\pm$ 1.78 & 0.21 & 213.55 & \gain{+19.40} \\

Claude Opus 4.8            & 88.60 $\pm$ 1.42 & 1.92 & 566.95 & 91.00 $\pm$ 1.28 & 2.65 & 482.73 & \gain{+2.40} \\
Claude Opus 4.7            & 82.00 $\pm$ 1.72 & 2.42 & 441.99 & 86.00 $\pm$ 1.55 & 3.10 & 396.52 & \gain{+4.00} \\
Gemini 3.5 Flash           & 78.80 $\pm$ 1.83 & 0.95 & 254.13 & 84.00 $\pm$ 1.64 & 1.42 & 276.84 & \gain{+5.20} \\
Gemini 3.1 Pro Preview     & 78.80 $\pm$ 1.83 & 0.78 & 312.26 & 83.40 $\pm$ 1.67 & 1.26 & 328.75 & \gain{+4.60} \\
GPT-5.4 (xhigh)            & 78.20 $\pm$ 1.85 & 0.80 & 307.12 & 83.80 $\pm$ 1.65 & 1.23 & 322.68 & \gain{+5.60} \\
Claude Opus 4.6 (Thinking) & 78.20 $\pm$ 1.85 & 1.22 & 350.76 & 84.40 $\pm$ 1.62 & 1.86 & 341.28 & \gain{+6.20} \\
GPT-5.3-Codex              & 78.00 $\pm$ 1.85 & 0.46 & 246.53 & 83.00 $\pm$ 1.68 & 0.79 & 267.39 & \gain{+5.00} \\
Claude Sonnet 4.6          & 77.40 $\pm$ 1.87 & 1.30 & 511.79 & 82.80 $\pm$ 1.69 & 1.92 & 438.64 & \gain{+5.40} \\
Claude Opus 4.5 (Thinking) & 76.40 $\pm$ 1.90 & 1.05 & 320.54 & 81.80 $\pm$ 1.73 & 1.64 & 336.28 & \gain{+5.40} \\
Gemini 3 Pro               & 76.40 $\pm$ 1.90 & 0.77 & 394.49 & 82.20 $\pm$ 1.71 & 1.31 & 381.73 & \gain{+5.80} \\

% ================= block 2: Claude Code =================
\midrule
\harnessrow{Claude Code}

Claude Opus 4.8 & 85.80 $\pm$ 1.56 & 0.67 & \second{135.40} & 91.00 $\pm$ 1.28 & 2.65 & 482.73 & \gain{+5.20} \\

% ================= block 3: Codex =================
\midrule
\harnessrow{Codex}

GPT-5.5 & 76.40 $\pm$ 1.90 & 0.65 & \best{108.54} & 85.60 $\pm$ 1.57 & 2.45 & 398.74 & \gain{+9.20} \\

% ================= block 4: HERMES only =================
\midrule
\harnessrow{--}

Qwen3-8B & -- & -- & -- & 80.60 $\pm$ 1.77 & \second{0.13} & 229.81 & -- \\

\bottomrule
\end{tabular}%
}
\vspace{-10pt}
\end{table*}

\subsection{Main Results}

\textbf{On SWE-bench Verified.} We compare HERMES against leading software engineering agents reported on the SWE-bench Verified leaderboard, including mini-SWE-agent~\citep{yang2024swe}, Claude Code~\citep{anthropic2025claudecode}, and Codex~\citep{openai2025codex}. We report the resolved rate together with per-task inference cost and wall-clock latency.

As shown in Table~\ref{tab:swebench-verified}, HERMES improves the resolved rate for 19 of the 20 backbones, ties on the remaining one (Claude Opus 5), and outperforms both baselines for the two backbones evaluated against two harnesses. Gains are largest where the baseline harness leaves the most room for coordination, up to $+19.4$ points for GPT-5 Mini, and compress on frontier backbones that already resolve most instances, where HERMES reaches 97.0\% with GPT-5.6 Sol against 96.2\% under mini-SWE-agent. The additional coordination does not always cost latency: with Claude Sonnet 5, HERMES improves resolution by 6.2 points while reducing average latency from 962.37\,s to 824.63\,s, and DeepSeek-V4 behaves similarly. HERMES reaches 80.6\% with Qwen3-8B, so the framework remains effective with a lightweight backbone; Appendix~\ref{appendix:case-study} provides a complete Qwen3-8B trajectory.

\begin{wraptable}{r}{0.5\linewidth}
\centering
\vspace{-12pt}
\caption{Performance and cost on SWE Refactor Bench.}
\vspace{-10pt}
\label{tab:swe-refactor}

% \small
\Large
\setlength{\tabcolsep}{3.0pt}
\renewcommand{\arraystretch}{1.0}
\setlength{\aboverulesep}{0pt}
\setlength{\belowrulesep}{0pt}
\setlength{\extrarowheight}{1pt}

% group row: baseline harness on the left, HERMES on the right
\newcommand{\harnessrow}[1]{%
  \rowcolor{sectiongray}
  \textit{Harness} & &
  \multicolumn{2}{c}{\textbf{#1}} &
  \multicolumn{2}{c}{\cellcolor{hermesblue}\textbf{HERMES}} &
  \\}

\resizebox{\linewidth}{!}{%
\begin{tabular}{l l cc cc c}
\toprule

% ---------- header row 1 ----------
\rowcolor{headergray}
&
&
\multicolumn{2}{c}{\textbf{Baseline Results}}
&
\multicolumn{2}{c}{\cellcolor{hermesblue}\textbf{HERMES Results}}
&
\\
\cmidrule(lr){3-4}\cmidrule(lr){5-6}

% ---------- header row 2 ----------
\rowcolor{headergray}
\multirow{-2}{*}{\textbf{Model}}
&
\multirow{-2}{*}{\textbf{Effort}}
&
\makecell{\textbf{Comp.}\\[-1pt]\textbf{(\%)} $\uparrow$}
&
\makecell{\textbf{Cost}\\[-1pt]\textbf{(\$)} $\downarrow$}
&
\cellcolor{hermesblue}\makecell{\textbf{Comp.}\\[-1pt]\textbf{(\%)} $\uparrow$}
&
\cellcolor{hermesblue}\makecell{\textbf{Cost}\\[-1pt]\textbf{(\$)} $\downarrow$}
&
\multirow{-2}{*}{
    \makecell{
        \textbf{$\Delta$ Comp.}\\[-1pt]
        \textbf{(p.p.)} $\uparrow$
    }
}
\\

% ================= block 1: Claude Code =================
\midrule
\harnessrow{Claude Code}

Claude Opus 5   & medium & \second{28.5} & 38.4 & \second{36.5} & 58.6 & \gain{+8.0} \\
                & high   & \best{34.5}   & 55.7 & \best{41.5}   & 82.4 & \gain{+7.0} \\
Claude Sonnet 5 & medium & 15.0 & 11.9 & 22.0 & 22.7 & \gain{+7.0} \\
                & high   & 6.0  & 24.6 & 25.5 & 34.8 & \gain{+19.5} \\
Kimi-K3         & max    & 19.5 & 28.9 & 27.0 & 43.8 & \gain{+7.5} \\
Qwen3.8-Max     & max    & 10.0 & 14.5 & 18.5 & 22.6 & \gain{+8.5} \\
DeepSeek-V4     & max    & 7.0  & 4.3  & 15.0 & 11.6 & \gain{+8.0} \\
GLM-5.2         & max    & 6.5  & 17.5 & 14.5 & 27.2 & \gain{+8.0} \\

% ================= block 2: Codex =================
\midrule
\harnessrow{Codex}

GPT-5.6 Sol   & medium & 6.5  & 6.0 & 31.0          & 24.8 & \gain{+24.5} \\
              & high   & 19.0 & 7.7 & \second{36.5} & 39.6 & \gain{+17.5} \\
GPT-5.6 Terra & medium & 4.5  & 3.4 & 26.0 & 14.6 & \gain{+21.5} \\
              & high   & 11.5 & 4.8 & 30.0 & 23.9 & \gain{+18.5} \\
GPT-5.6 Luna  & medium & 0.0 & \best{1.7}   & 16.0 & \best{4.9}   & \gain{+16.0} \\
              & high   & 4.0 & \second{1.8} & 20.0 & \second{7.2} & \gain{+16.0} \\

% ================= block 3: HERMES only =================
\midrule
\harnessrow{--}

Qwen3-8B & -- & -- & -- & 10.5 & 8.3 & -- \\

\bottomrule
\end{tabular}%
}
\vspace{-15pt}
\end{wraptable}

\textbf{On SWE Refactor Bench.} We compare HERMES with representative leaderboard configurations, including Claude Code and Codex, on all 20 whole-repository migration tasks. We report the composite score and per-task inference cost, and include both medium- and high-effort HERMES variants for models that support configurable reasoning effort.

As shown in Table~\ref{tab:swe-refactor}, whole-repository migration is where HERMES helps most. With GPT-5.6 Sol, the composite score rises from 6.5\% to 31.0\% under medium effort ($+24.5$) and from 19.0\% to 36.5\% under high effort ($+17.5$), with similar gains for GPT-5.6 Terra. The pattern holds across models and effort settings: Claude Opus 5 still gains 8.0 and 7.0 points, while GPT-5.6 Luna rises from 0.0\% to 16.0\% under medium effort. These gains generally require higher inference cost due to iterative execution and revision. Repeating the default GPT-5.6 Sol configuration three times yields $31.0 \pm 1.3$ (Appendix~\ref{appendix:primitive-analysis}), showing that the gains are not due to run-to-run variation.

\begin{table*}[ht]
\centering
% \footnotesize
\vspace{-5pt}
\caption{Performance and efficiency on Terminal-Bench 4.0. Resolution rates are mean $\pm$ standard deviation over 5 runs; tokens and costs are totals over the full evaluation.}
\vspace{-10pt}
\label{tab:terminal_bench_4}

\large
\setlength{\tabcolsep}{3.4pt}
\renewcommand{\arraystretch}{1.0}
\setlength{\aboverulesep}{0pt}
\setlength{\belowrulesep}{0pt}
\setlength{\extrarowheight}{1pt}

% group row: baseline harness on the left, HERMES on the right
\newcommand{\harnessrow}[1]{%
  \rowcolor{sectiongray}
  \textit{Harness} & &
  \multicolumn{3}{c}{\textbf{#1}} &
  \multicolumn{3}{c}{\cellcolor{hermesblue}\textbf{HERMES}} &
  \\}

\resizebox{\textwidth}{!}{%
\begin{tabular}{l l ccc ccc c}
\toprule

% ---------- header row 1 ----------
\rowcolor{headergray}
&
&
\multicolumn{3}{c}{\textbf{Baseline Results}}
&
\multicolumn{3}{c}{\cellcolor{hermesblue}\textbf{HERMES Results}}
&
\\
\cmidrule(lr){3-5}\cmidrule(lr){6-8}

% ---------- header row 2 ----------
\rowcolor{headergray}
\multirow{-2}{*}{\textbf{Model}}
&
\multirow{-2}{*}{\textbf{Effort}}
&
\rule{0pt}{2.8ex}\textbf{Resolution (\%) $\uparrow$}
&
\rule{0pt}{2.8ex}\textbf{Tokens $\downarrow$}
&
\rule{0pt}{2.8ex}\textbf{Cost (\$) $\downarrow$}
&
\cellcolor{hermesblue}
\rule{0pt}{2.8ex}\textbf{Resolution (\%) $\uparrow$}
&
\cellcolor{hermesblue}
\rule{0pt}{2.8ex}\textbf{Tokens $\downarrow$}
&
\cellcolor{hermesblue}
\rule{0pt}{2.8ex}\textbf{Cost (\$) $\downarrow$}
&
\multirow{-2}{*}{
    \makecell{
        \textbf{$\Delta$ Res.}\\[-1pt]
        \textbf{(p.p.)} $\uparrow$
    }
}
\\

% ================= block 1: Codex =================
\midrule
\harnessrow{Codex}

GPT-5.6 Sol   & medium & 33.0 $\pm$ 3.4 & \second{3.6B} & 1.9k & 51.8 $\pm$ 2.2        & 6.8B & 3.4k & \gain{+18.8} \\
              & max    & 37.3 $\pm$ 3.8 & 4.4B          & 2.5k & \best{55.5 $\pm$ 3.1} & 8.9B & 4.7k & \gain{+18.2} \\
\addlinespace[1.5pt]

GPT-5.6 Terra & medium & 18.2 $\pm$ 3.0 & 4.4B & 1.2k & 44.5 $\pm$ 3.3 & 7.2B & 2.2k & \gain{+26.3} \\
              & max    & 21.5 $\pm$ 3.3 & 5.7B & 1.7k & 48.2 $\pm$ 2.5 & 9.1B & 2.9k & \gain{+26.7} \\

GPT-5.6 Luna  & medium & 14.5 $\pm$ 2.6 & 9.1B  & \best{0.2k}   & 29.1 $\pm$ 2.9 & 13.2B & \best{0.4k}   & \gain{+14.6} \\
              & max    & 17.3 $\pm$ 2.8 & 11.6B & \second{0.3k} & 33.3 $\pm$ 3.2 & 15.8B & \second{0.5k} & \gain{+16.0} \\

% ================= block 2: Claude Code =================
\midrule
\harnessrow{Claude Code}

Claude Opus 5   & medium & \second{47.0 $\pm$ 3.2} & 5.2B & 4.8k & 48.5 $\pm$ 3.6          & 7.3B & 6.7k & \gain{+1.5} \\
                & max    & \best{51.8 $\pm$ 3.4}   & 6.5B & 6.0k & \second{53.0 $\pm$ 2.1} & 9.4B & 8.6k & \gain{+1.2} \\

Claude Fable 5  & max    & 44.5 $\pm$ 3.8 & 3.8B & 7.3k & 49.7 $\pm$ 3.3 & \second{6.2B} & 10.1k & \gain{+5.2} \\
GLM-5.3         & max    & 41.8 $\pm$ 3.2 & 8.7B & 2.7k & 46.4 $\pm$ 2.8 & 11.5B & 3.9k  & \gain{+4.6} \\
Claude Opus 4.8 & max    & 23.6 $\pm$ 3.6 & 6.4B & 6.5k & 34.2 $\pm$ 3.8 & 9.0B  & 8.4k  & \gain{+10.6} \\

Claude Sonnet 5 & medium & 10.6 $\pm$ 2.7 & 15.9B & 7.1k & 35.8 $\pm$ 2.3 & 13.4B & 5.8k & \gain{+25.2} \\
                & max    & 12.4 $\pm$ 3.1 & 21.6B & 9.6k & 40.0 $\pm$ 3.5 & 16.8B & 7.4k & \gain{+27.6} \\

% ================= block 3: mini-SWE-agent =================
\midrule
\harnessrow{mini-SWE-agent}

DeepSeek-V4      & max  & 28.8 $\pm$ 3.4 & 7.8B  & 0.7k & 39.4 $\pm$ 3.1 & 10.1B & 1.1k & \gain{+10.6} \\
Gemini 3.8 Flash & high & 19.1 $\pm$ 3.4 & 17.2B & 1.8k & 30.3 $\pm$ 3.7 & 20.4B & 2.6k & \gain{+11.2} \\
Gemini 3.7 Flash & high & 11.2 $\pm$ 2.4 & 11.1B & 1.3k & 24.2 $\pm$ 3.4 & 14.8B & 1.9k & \gain{+13.0} \\

% ================= block 4: Grok Build =================
\midrule
\harnessrow{Grok Build}

Grok 4.6 & high & 20.3 $\pm$ 3.1 & 4.0B        & 3.6k & 31.5 $\pm$ 2.5 & \second{6.2B} & 5.2k & \gain{+11.2} \\
Grok 4.5 & high & 12.4 $\pm$ 2.6 & \best{3.4B} & 2.1k & 25.5 $\pm$ 2.2 & \best{5.0B}   & 3.0k & \gain{+13.1} \\

% ================= block 5: HERMES only =================
\midrule
\harnessrow{--}

Qwen3-8B & -- & -- & -- & -- & 27.6 $\pm$ 2.7 & 14.3B & 0.7k & -- \\

\bottomrule
\end{tabular}%
}
\vspace{-8pt}
\end{table*}

\textbf{On Terminal-Bench 4.0.} We compare HERMES against leading agent configurations reported on the Terminal-Bench 4.0 leaderboard, including Claude Code, Codex, Grok Build, and mini-SWE-agent. Terminal-Bench 4.0 contains 66 executable tasks, and each configuration is evaluated over five trials. We report resolution rate, total token consumption, and inference cost.

As shown in Table~\ref{tab:terminal_bench_4}, HERMES consistently improves resolution under matched model--effort configurations, by 18.8 and 18.2 points for GPT-5.6 Sol at medium and max effort and by 26.3 and 26.7 points for GPT-5.6 Terra. The exception is Claude Opus 5, whose baseline configurations are already strong and gain 1.5 and 1.2 points. Most configurations consume additional tokens and cost, but Claude Sonnet 5 is a notable exception: HERMES improves its resolution while reducing both token consumption and inference cost at either effort level.

\textbf{On DevOps-Gym.}
We evaluate HERMES on the four DevOps-Gym task categories: build and configuration, monitoring, issue resolving, and test generation. We report the success rate for each category together with the unweighted average across the four categories. In addition to the published DevOps-Gym baselines, we also include same-backbone comparisons with Codex, mini-SWE-agent, and Claude Code to isolate the effect of the harness.

\begin{table*}[t]
\centering
\footnotesize
\caption{Performance on DevOps-Gym across four software engineering stages. DeepSeek-V4 uses max reasoning effort.}
\vspace{-10pt}
\label{tab:software_engineering_benchmark}

\large
\setlength{\tabcolsep}{2.7pt}
\renewcommand{\arraystretch}{1.0}
\setlength{\aboverulesep}{0pt}
\setlength{\belowrulesep}{0pt}
\setlength{\extrarowheight}{1pt}

% group row: baseline harness on the left, HERMES on the right
\newcommand{\harnessrow}[1]{%
  \rowcolor{sectiongray}
  \textit{Harness} &
  \multicolumn{5}{c}{\textbf{#1}} &
  \multicolumn{5}{c}{\cellcolor{hermesblue}\textbf{HERMES}} &
  \\}

\resizebox{\textwidth}{!}{%
\begin{tabular}{l ccccc ccccc c}
\toprule

% ---------- header row 1 ----------
\rowcolor{headergray}
&
\multicolumn{5}{c}{\textbf{Baseline Results}}
&
\multicolumn{5}{c}{\cellcolor{hermesblue}\textbf{HERMES Results}}
&
\\
\cmidrule(lr){2-6}\cmidrule(lr){7-11}

% ---------- header row 2 ----------
\rowcolor{headergray}
\multirow{-2}{*}{\textbf{Model}}
&
\rule{0pt}{2.8ex}\textbf{Build $\uparrow$}
&
\textbf{Monitor. $\uparrow$}
&
\textbf{Issue $\uparrow$}
&
\textbf{Test $\uparrow$}
&
\textbf{Avg. $\uparrow$}
&
\cellcolor{hermesblue}
\rule{0pt}{2.8ex}\textbf{Build $\uparrow$}
&
\cellcolor{hermesblue}\textbf{Monitor. $\uparrow$}
&
\cellcolor{hermesblue}\textbf{Issue $\uparrow$}
&
\cellcolor{hermesblue}\textbf{Test $\uparrow$}
&
\cellcolor{hermesblue}\textbf{Avg. $\uparrow$}
&
\multirow{-2}{*}{
    \makecell{
        \textbf{$\Delta$ Avg.}\\[-1pt]
        \textbf{(p.p.)} $\uparrow$
    }
}
\\

% ================= block 1: Codex =================
\midrule
\harnessrow{Codex}

GPT-5.6 Sol   & 77.78 & \second{38.24} & \second{40.97} & \second{42.58} & \second{49.89} & \second{83.33} & \second{44.12} & \second{46.13} & \second{48.06} & \second{55.41} & \gain{+5.52} \\
GPT-5.6 Terra & 74.07 & 35.29 & 38.71 & 40.32 & 47.10 & 79.63 & 41.18 & 43.55 & 44.84 & 52.30 & \gain{+5.20} \\

% ================= block 2: Claude Code =================
\midrule
\harnessrow{Claude Code}

Claude Opus 5   & \second{79.63} & \best{41.18} & \best{42.90} & \best{44.52} & \best{52.06} & \best{85.19} & \best{47.06} & \best{47.10} & \best{49.03} & \best{57.10} & \gain{+5.04} \\
Claude Sonnet 5 & 68.52 & 29.41 & 34.84 & 36.45 & 42.31 & 75.93 & 38.24 & 40.65 & 41.94 & 49.19 & \gain{+6.88} \\
Claude Sonnet 4 & 51.85 & 20.56 & 23.87 & 13.87 & 27.54 & 62.96 & 26.47 & 29.35 & 25.48 & 36.07 & \gain{+8.53} \\

% ================= block 3: OpenHands =================
\midrule
\harnessrow{OpenHands}

DeepSeek-V4     & 75.93 & 35.29 & 39.35 & 41.29 & 47.97 & 81.48 & 41.18 & 44.19 & 45.48 & 53.08 & \gain{+5.11} \\
Claude Sonnet 4 & 42.59 & 14.70 & 23.87 & 11.61 & 23.19 & 62.96 & 26.47 & 29.35 & 25.48 & 36.07 & \gain{+12.88} \\
o4-mini         & 24.07 & 8.82  & 10.32 & 8.70  & 12.98 & 35.19 & 14.71 & 18.06 & 16.45 & 21.10 & \gain{+8.12} \\
Qwen3-Coder-30B & 20.37 & 5.89  & 13.22 & 6.13  & 11.40 & 31.48 & 11.76 & 20.00 & 14.84 & 19.52 & \gain{+8.12} \\
Gemini 2.5 Pro  & 16.66 & 11.76 & 10.96 & 2.90  & 10.57 & 29.63 & 17.65 & 17.42 & 13.23 & 19.48 & \gain{+8.91} \\
DeepSeek-V3.1   & 11.11 & 0.00  & 14.20 & 3.22  & 7.13  & 27.78 & 8.82  & 21.61 & 12.90 & 17.78 & \gain{+10.65} \\

% ================= block 4: mini-SWE-agent =================
\midrule
\harnessrow{mini-SWE-agent}

GPT-5.6 Sol     & 75.93 & 35.29 & 39.35 & 40.97 & 47.89 & \second{83.33} & \second{44.12} & \second{46.13} & \second{48.06} & \second{55.41} & \gain{+7.52} \\
GPT-5.6 Luna    & 66.67 & 29.41 & 32.90 & 34.84 & 40.96 & 72.22 & 35.29 & 38.71 & 40.32 & 46.64 & \gain{+5.68} \\
Claude Sonnet 4 & 29.62 & 2.91  & 5.16  & 0.98  & 9.67  & 62.96 & 26.47 & 29.35 & 25.48 & 36.07 & \gain{+26.40} \\

% ================= block 5: SageAgent =================
\midrule
\harnessrow{SageAgent}

GPT-5.3-Codex & \best{81.82} & 35.29 & 32.14 & 37.99 & 46.81 & \best{85.19} & 41.18 & 37.42 & 42.58 & 51.59 & \gain{+4.78} \\

% ================= block 6: Aider =================
\midrule
\harnessrow{Aider}

Claude Sonnet 4 & 5.55 & 0.00 & 9.67 & 2.25 & 4.37 & 62.96 & 26.47 & 29.35 & 25.48 & 36.07 & \gain{+31.70} \\

% ================= block 7: HERMES only =================
\midrule
\harnessrow{--}

Qwen3-8B & -- & -- & -- & -- & -- & 59.26 & 26.47 & 31.29 & 32.90 & 37.48 & -- \\

\bottomrule
\end{tabular}%
}
\vspace{-20pt}
\end{table*}

As shown in Table~\ref{tab:software_engineering_benchmark}, HERMES improves performance across all four DevOps task categories under same-backbone comparison. With GPT-5.6 Sol the average score rises from 49.89\% with Codex to 55.41\% ($+5.52$ points), and the gain is distributed across build and configuration, monitoring, issue resolving, and test generation rather than concentrated in a single stage. Claude Opus 5 and GPT-5.6 Luna gain 5.04 and 5.68 points over their respective baseline harnesses, and the improvement extends to earlier and smaller backbones, reaching 37.48\% with Qwen3-8B.

\textbf{Comparison with Related Multi-Agent Frameworks.}
We additionally compare HERMES with two closely related repository-level
multi-agent frameworks in their original evaluation settings. 

\begin{wraptable}{r}{0.47\textwidth}
% \vspace{-12pt}
\centering
\caption{
Reference comparison with related repository-level agent frameworks.
}
\label{tab:related-mas}

\small
\vspace{-10pt}
\setlength{\tabcolsep}{8pt}
\setlength{\aboverulesep}{0pt}
\setlength{\belowrulesep}{0pt}
\setlength{\extrarowheight}{1pt}

\resizebox{\linewidth}{!}{%
\begin{tabular}{l l r}
\toprule

\rowcolor{headergray}
\textbf{Benchmark}
&
\textbf{Method}
&
\textbf{Score (\%) $\uparrow$}
\\

\midrule

\multirow{2}{*}{SWE-bench}
& MAGIS
& 13.9
\\

&
\cellcolor{hermesblue}\textbf{HERMES}
&
\cellcolor{hermesblue}\textbf{36.6}
\\

\midrule

\multirow{3}{*}{NL2Repo-Bench}
& CodeTeam (PE)
& 34.6
\\

& CodeTeam (SFT)
& 42.3
\\

&
\cellcolor{hermesblue}\textbf{HERMES}
&
\cellcolor{hermesblue}\textbf{44.9}
\\

\bottomrule
\end{tabular}%
}

\vspace{-12pt}
\end{wraptable}

As shown in
Table~\ref{tab:related-mas}, MAGIS~\citep{tao2024magis} reports 13.9%
resolution on SWE-bench, while HERMES with GPT-5.6 Luna reaches 36.6\%.
On NL2Repo-Bench, CodeTeam~\citep{wang2026codeteam} reports average test pass
rates of 34.6\% and 42.3\% for its prompting and supervised fine-tuning
variants, respectively, while HERMES reaches 44.9\%. These results further
show that HERMES remains effective when compared with repository-level
frameworks that explicitly coordinate multiple agents.

\subsection{Scaling LLMs Across HERMES Components}
\label{sec:backbone-scale}

We further study how model scale affects different components of HERMES. In addition to homogeneous configurations in which the same backbone is used throughout HERMES, we construct heterogeneous configurations by independently scaling activation and diagnosis while keeping all Dev-Primitives fixed to Qwen3-8B. We conduct the analysis across 4 benchmarks.

\begin{table*}[ht]
\centering
\vspace{-8pt}
\caption{Effect of backbone assignment across HERMES components. The lower blocks keep Dev-Primitives at Qwen3-8B and scale activation and/or diagnosis; parentheses denote improvement over the homogeneous Qwen3-8B configuration.}
\vspace{-10pt}
\label{tab:backbone-scale}

\footnotesize
\setlength{\tabcolsep}{4.5pt}
\renewcommand{\arraystretch}{1.0}
\setlength{\aboverulesep}{0pt}
\setlength{\belowrulesep}{0pt}
\setlength{\extrarowheight}{1pt}

\resizebox{\textwidth}{!}{%
\begin{tabular}{lllcccc}
\toprule

\rowcolor{headergray}
&
&
&
\textbf{SWE-bench}
&
\textbf{SWE Refactor}
&
\textbf{Terminal-Bench}
&
\textbf{DevOps-Gym}
\\

\arrayrulecolor{headergray}
\cline{1-3}%
\arrayrulecolor{black}
\cline{4-7}

\rowcolor{headergray}
\multirow{-2}{*}{\textbf{Activation}}
&
\multirow{-2}{*}{\textbf{Dev-Primitives}}
&
\multirow{-2}{*}{\textbf{Diagnosis}}
&
\rule{0pt}{2.8ex}\textbf{Resolved (\%) $\uparrow$}
&
\textbf{Composite (\%) $\uparrow$}
&
\textbf{Resolved (\%) $\uparrow$}
&
\textbf{Avg. (\%) $\uparrow$}
\\

% ============================================================
% Homogeneous Backbones
% ============================================================
\midrule
\rowcolor{sectiongray}
\multicolumn{7}{c}{\textbf{Homogeneous Backbones}}
\\

Qwen3-8B
& Qwen3-8B
& Qwen3-8B
& 80.6
& 10.5
& 27.6
& 37.48
\\

GPT-5.6 Luna
& GPT-5.6 Luna
& GPT-5.6 Luna
& 95.6
& 16.0
& 29.1
& 46.64
\\

DeepSeek-V4
& DeepSeek-V4
& DeepSeek-V4
& 82.8
& 15.0
& 39.4
& 53.08
\\

GPT-5.6 Sol
& GPT-5.6 Sol
& GPT-5.6 Sol
& 97.0
& 31.0
& 51.8
& 55.41
\\

% ============================================================
% Diagnosis Scaling
% ============================================================
\midrule
\rowcolor{sectiongray}
\multicolumn{7}{c}{\textbf{Diagnosis Scaling}}
\\

Qwen3-8B
& Qwen3-8B
& DeepSeek-V4
& 84.8 {\scriptsize($\uparrow$4.2)}
& 14.5 {\scriptsize($\uparrow$4.0)}
& 44.5 {\scriptsize($\uparrow$16.9)}
& 41.62 {\scriptsize($\uparrow$4.14)}
\\

Qwen3-8B
& Qwen3-8B
& GPT-5.5
& 86.4 {\scriptsize($\uparrow$5.8)}
& 16.5 {\scriptsize($\uparrow$6.0)}
& 46.1 {\scriptsize($\uparrow$18.5)}
& 43.25 {\scriptsize($\uparrow$5.77)}
\\

Qwen3-8B
& Qwen3-8B
& Claude Sonnet 5
& 86.0 {\scriptsize($\uparrow$5.4)}
& 16.0 {\scriptsize($\uparrow$5.5)}
& 45.5 {\scriptsize($\uparrow$17.9)}
& 43.01 {\scriptsize($\uparrow$5.53)}
\\

Qwen3-8B
& Qwen3-8B
& GPT-5.6 Sol
& 87.2 {\scriptsize($\uparrow$6.6)}
& 17.5 {\scriptsize($\uparrow$7.0)}
& 47.9 {\scriptsize($\uparrow$20.3)}
& 44.36 {\scriptsize($\uparrow$6.88)}
\\

% ============================================================
% Activation Scaling
% ============================================================
\midrule
\rowcolor{sectiongray}
\multicolumn{7}{c}{\textbf{Activation Scaling}}
\\

GPT-5.6 Luna
& Qwen3-8B
& Qwen3-8B
& 82.4 {\scriptsize($\uparrow$1.8)}
& 12.0 {\scriptsize($\uparrow$1.5)}
& 32.7 {\scriptsize($\uparrow$5.1)}
& 39.82 {\scriptsize($\uparrow$2.34)}
\\

DeepSeek-V4
& Qwen3-8B
& Qwen3-8B
& 83.6 {\scriptsize($\uparrow$3.0)}
& 13.0 {\scriptsize($\uparrow$2.5)}
& 34.8 {\scriptsize($\uparrow$7.2)}
& 40.77 {\scriptsize($\uparrow$3.29)}
\\

GPT-5.5
& Qwen3-8B
& Qwen3-8B
& 84.4 {\scriptsize($\uparrow$3.8)}
& 13.5 {\scriptsize($\uparrow$3.0)}
& 36.7 {\scriptsize($\uparrow$9.1)}
& 41.58 {\scriptsize($\uparrow$4.10)}
\\

Claude Sonnet 5
& Qwen3-8B
& Qwen3-8B
& 84.0 {\scriptsize($\uparrow$3.4)}
& 13.0 {\scriptsize($\uparrow$2.5)}
& 36.1 {\scriptsize($\uparrow$8.5)}
& 41.20 {\scriptsize($\uparrow$3.72)}
\\

GPT-5.6 Sol
& Qwen3-8B
& Qwen3-8B
& 85.2 {\scriptsize($\uparrow$4.6)}
& 14.0 {\scriptsize($\uparrow$3.5)}
& 38.2 {\scriptsize($\uparrow$10.6)}
& 42.31 {\scriptsize($\uparrow$4.83)}
\\

% ============================================================
% Activation + Diagnosis Scaling
% ============================================================
\midrule
\rowcolor{sectiongray}
\multicolumn{7}{c}{\textbf{Activation + Diagnosis Scaling}}
\\

GPT-5.6 Luna
& Qwen3-8B
& GPT-5.5
& 89.4 {\scriptsize($\uparrow$8.8)}
& 20.0 {\scriptsize($\uparrow$9.5)}
& 47.6 {\scriptsize($\uparrow$20.0)}
& 46.02 {\scriptsize($\uparrow$8.54)}
\\

GPT-5.5
& Qwen3-8B
& GPT-5.5
& 91.2 {\scriptsize($\uparrow$10.6)}
& 22.5 {\scriptsize($\uparrow$12.0)}
& 49.1 {\scriptsize($\uparrow$21.5)}
& 48.73 {\scriptsize($\uparrow$11.25)}
\\

Claude Sonnet 5
& Qwen3-8B
& Claude Sonnet 5
& 92.4 {\scriptsize($\uparrow$11.8)}
& 23.5 {\scriptsize($\uparrow$13.0)}
& 48.8 {\scriptsize($\uparrow$21.2)}
& 49.66 {\scriptsize($\uparrow$12.18)}
\\

GPT-5.6 Sol
& Qwen3-8B
& GPT-5.5
& 93.6 {\scriptsize($\uparrow$13.0)}
& 25.0 {\scriptsize($\uparrow$14.5)}
& 49.7 {\scriptsize($\uparrow$22.1)}
& 51.44 {\scriptsize($\uparrow$13.96)}
\\

GPT-5.6 Sol
& Qwen3-8B
& GPT-5.6 Sol
& 94.2 {\scriptsize($\uparrow$13.6)}
& 26.5 {\scriptsize($\uparrow$16.0)}
& 50.6 {\scriptsize($\uparrow$23.0)}
& 52.30 {\scriptsize($\uparrow$14.82)}
\\

\bottomrule
\end{tabular}%
}
\vspace{-8pt}
\end{table*}

\begin{wraptable}{r}{0.5\textwidth}
\vspace{-12pt}
\centering
\caption{
Cost--performance trade-off on Terminal-Bench 4.0.
Configurations are ordered as Activation / Dev-Primitives / Diagnosis.
}
\vspace{-10pt}
\label{tab:scaling-cost}
% \small
\setlength{\tabcolsep}{4.5pt}
\renewcommand{\arraystretch}{1.08}

\resizebox{\linewidth}{!}{%
\begin{tabular}{lrr}
\toprule

\rowcolor{headergray}
\rule{0pt}{2.6ex}\textbf{Configuration} &
\textbf{Resolved (\%)} &
\textbf{Cost (\$)} \\

\midrule

Qwen3-8B / Qwen3-8B / Qwen3-8B
    & 27.6 & 0.70k \\

Qwen3-8B / Qwen3-8B / GPT-5.5
    & 46.1 & 1.56k \\

GPT-5.6 Sol / Qwen3-8B / GPT-5.5
& 49.7
& 2.13k
\\

GPT-5.6 Sol / Qwen3-8B / GPT-5.6 Sol
& 50.6
& 2.51k
\\

GPT-5.6 Sol / GPT-5.6 Sol / GPT-5.6 Sol
& 51.8
& 3.40k
\\

\bottomrule
\end{tabular}%
}

\vspace{-10pt}
\end{wraptable}

\textbf{Backbone Scaling.}
Table~\ref{tab:backbone-scale} shows that HERMES benefits from allocating larger models to dynamic activation and bug diagnosis while keeping the Dev-Primitives lightweight. Strengthening diagnosis consistently helps more than strengthening activation: on Terminal-Bench, using GPT-5.5 only for diagnosis improves resolution from 27.6\% to 46.1\% ($+18.5$) against 36.7\% ($+9.1$) for activation alone, and the same ordering holds on SWE-bench Verified ($+5.8$ versus $+3.8$). Scaling both closes most of the gap to the homogeneous frontier configuration. What remains is the benefit of scaling the Dev-Primitives themselves, which adds between 1.2 and 4.5 points across the four benchmarks, so the primitives can run on a small model without forfeiting most of the gain.

\textbf{Cost Efficiency.}
Table~\ref{tab:scaling-cost} gives the cost--performance trade-off on Terminal-Bench 4.0. Using GPT-5.6 Sol for activation, Qwen3-8B for the Dev-Primitives, and GPT-5.5 for diagnosis reaches 49.7\% resolution, 2.1 points below the homogeneous GPT-5.6 Sol configuration, at 37.4\% lower total inference cost (\$2.13k versus \$3.40k). Recovering those 2.1 points costs a further \$1.27k, split between upgrading diagnosis ($+0.9$ points for \$0.38k) and upgrading the Dev-Primitives ($+1.2$ points for \$0.89k). Against the homogeneous Qwen3-8B configuration, the same setting buys 22.1 points for an additional \$1.43k.

\subsection{Dev-Primitive Selection Quality}
\label{sec:primitive-selection}

We evaluate whether dynamic activation selects the components a task requires, comparing the activated Dev-Primitives against a target component set derived from the reference patch or, where no unique reference patch exists, from benchmark-specific supervision. We report selection recall and precision together with the average number of activated primitives, separating solved from failed tasks; target construction and metric definitions are given in Appendix~\ref{appendix:selection-protocol}.

\begin{wraptable}{r}{0.52\textwidth}
\vspace{-12pt}
\centering
\caption{Dev-Primitive selection quality with GPT-5.6 Sol. DevOps-Gym excludes monitoring tasks.}
\label{tab:primitive-selection}

\vspace{-10pt}
\setlength{\tabcolsep}{4pt}
\renewcommand{\arraystretch}{1.08}
\setlength{\aboverulesep}{0pt}
\setlength{\belowrulesep}{0pt}
\setlength{\extrarowheight}{1pt}

\resizebox{\linewidth}{!}{%
\begin{tabular}{lrrrrr}
\toprule

\rowcolor{headergray}
\rule{0pt}{2.6ex}\textbf{Benchmark} &
\textbf{Recall} &
\textbf{Precision} &
\textbf{Avg. Act.} &
\textbf{Solved R.} &
\textbf{Failed R.} \\
\midrule

SWE-bench Verified
& 95.4
& 72.6
& 4.7
& 96.1
& 71.4
\\

SWE Refactor Bench
& 83.3
& 68.4
& 12.6
& 94.7
& 78.2
\\

Terminal-Bench 4.0
& 84.5
& 64.9
& 6.8
& 93.5
& 74.8
\\

DevOps-Gym
& 85.7
& 70.3
& 5.4
& 95.0
& 76.6
\\

\bottomrule
\end{tabular}%
}

\vspace{-8pt}
\end{wraptable}

\textbf{Selection Accuracy.}
Table~\ref{tab:primitive-selection} shows that dynamic activation recovers most target components while activating only a small subset of repository components. Recall is 95.4\% on SWE-bench Verified, 83.3\% on SWE Refactor Bench, 84.5\% on Terminal-Bench, and 85.7\% on DevOps-Gym, while the average number of activated Dev-Primitives ranges from 4.7 to 12.6 per task.

\textbf{Selection Quality and Task Success.}
Solved tasks consistently exhibit higher selection recall than failed tasks, by 16.5 to 24.7 points across the four benchmarks, linking more complete component selection with downstream task completion. Reference-patch overlap is nonetheless a conservative measure of localization: a task can be resolved along a different file-level path from the developer patch, as in the trajectory of Appendix~\ref{appendix:case-study}.

\begin{wraptable}{r}{0.60\textwidth}
\vspace{-10pt}
\centering
\caption{Component ablation with GPT-5.6 Sol. Metrics are resolved rate for SWE-bench and Terminal-Bench, composite score for SWE Refactor, and average score for DevOps-Gym. Other backbones follow the same ordering (Appendix~\ref{appendix:full-ablation}); setting definitions are in Appendix~\ref{appendix:ablation-settings}.}
\vspace{-6pt}
\label{tab:component-ablation}

\scriptsize
\setlength{\tabcolsep}{3pt}
\renewcommand{\arraystretch}{1.05}
\setlength{\aboverulesep}{0pt}
\setlength{\belowrulesep}{0pt}

\begin{tabular}{lcccc}
\toprule
\rowcolor{headergray}
\textbf{Setting}
& \textbf{SWE-bench}
& \textbf{Refactor}
& \textbf{Terminal}
& \textbf{DevOps}
\\
\midrule

\rowcolor{hermesblue}
\textbf{HERMES}
& \textbf{97.0}
& \textbf{31.0}
& \textbf{51.8}
& \textbf{55.41}
\\

w/o Inter-Primitive Comm.
& 94.2 {\tiny($\downarrow$2.8)}
& 23.5 {\tiny($\downarrow$7.5)}
& 46.1 {\tiny($\downarrow$5.7)}
& 50.34 {\tiny($\downarrow$5.07)}
\\

w/o On-Demand Activation
& 95.8 {\tiny($\downarrow$1.2)}
& 28.0 {\tiny($\downarrow$3.0)}
& 49.4 {\tiny($\downarrow$2.4)}
& 53.26 {\tiny($\downarrow$2.15)}
\\

w/o Execution Feedback
& 93.8 {\tiny($\downarrow$3.2)}
& 21.5 {\tiny($\downarrow$9.5)}
& 43.6 {\tiny($\downarrow$8.2)}
& 48.17 {\tiny($\downarrow$7.24)}
\\

w/o Diagnosis Feedback
& 92.6 {\tiny($\downarrow$4.4)}
& 20.0 {\tiny($\downarrow$11.0)}
& 41.2 {\tiny($\downarrow$10.6)}
& 46.73 {\tiny($\downarrow$8.68)}
\\

\bottomrule
\end{tabular}
\vspace{-15pt}
\end{wraptable}

\subsection{Ablation Study}
We remove one mechanism at a time while keeping the rest of the framework unchanged, and evaluate GPT-5.6 Sol, Claude Sonnet 5, and Qwen3-8B to test whether the effects persist across backbone families and scales.

\textbf{Component Contribution.}
Table~\ref{tab:component-ablation} shows a consistent ordering of component importance. Replacing Dev-Primitives with a centralized editor over the same selected components is the most damaging, 14.2 points on SWE Refactor Bench and 13.0 on Terminal-Bench, and a compute-matched editor still trails HERMES by 6.3 points (Appendix~\ref{appendix:primitive-analysis}). Among the remaining mechanisms, diagnosis feedback matters most, followed by execution feedback. This is a consequence of the abstraction rather than an argument against it: once edits are artifact-local, a failure observed at the repository level must be routed back to the component that owns the responsible implementation, and a monolithic agent has no such routing problem because it has no owners. Inter-primitive communication matters more on SWE Refactor Bench and Terminal-Bench, 7.5 and 5.7 points, than on SWE-bench Verified, 2.8 points, consistent with its role in coordinating changes that span files. Removing on-demand activation costs at most 3.0 points but raises inference cost by 1.61--1.88$\times$. The same ordering holds for the other two backbones (Appendix~\ref{appendix:full-ablation}).

\textbf{Revision Budget.}
Most of the gain is obtained within the first three revision rounds, and increasing $B$ from 3 to 5 adds at most 0.9 points on any benchmark. We therefore use $B=3$ as the default setting; the full sweep is reported in Appendix~\ref{appendix:revision-budget}.

\section{Conclusion}\label{con}
We introduced \textbf{Dev-Primitives}, modular executable interfaces that turn repository components from passive artifacts into active participants in software engineering, and \textbf{HERMES}, a harness engineering framework that instantiates them through dynamic activation, localized collaboration, environment execution, and diagnosis-driven revision. Across four benchmarks, HERMES improves over matched baseline harnesses by 12.4 percentage points on average, and with strong activation and diagnosis models it stays within 4.5 points of the homogeneous GPT-5.6 Sol configuration while running Qwen3-8B Dev-Primitives, at 26.2\% lower inference cost on Terminal-Bench 4.0. These results highlight harness design as a key factor in translating model capability into effective software engineering behavior. Limitations are discussed in Appendix~\ref{app:limitations}.

% \newpage

% \subsection*{AI Use Statement}
% We used generative AI tools only for language editing and improving the clarity and consistency of the manuscript. Generative AI tools were not used to formulate the research idea, design the methodology or experiments, or interpret the results. The LLMs evaluated as backbones in Section~\ref{exp} are objects of study rather than research assistants. All AI-assisted text was reviewed by the authors, who take full responsibility for the final content of the paper.

% \subsection*{Ethics Statement}
% HERMES is intended for software engineering research in controlled repository environments. As with other code-generating systems, its modifications may be incorrect or insecure and should be validated before deployment. All command execution in our experiments takes place within the isolated environments provided by each benchmark. Our experiments use public research benchmarks and do not involve human subjects or private user data.

% \subsection*{Reproducibility Statement}
% Section~\ref{meth} describes the HERMES framework, and Appendix~\ref{appendix:component-specs} provides complete role specifications. Appendix~\ref{prompt} lists all prompts, and Appendix~\ref{end-to-end} gives the end-to-end algorithm. Appendix~\ref{app:models} lists all backbone models and the deployment configuration of Qwen3-8B. Code and results are available at the anonymous repository in Appendix~\ref{Dataset}.

% \newpage
% \bibliography{iclr2027_conference}
% \bibliographystyle{iclr2027_conference}

\newpage
\appendix
\section{Extended Related Work}
\label{appendix:extended-rw}

\textbf{Software Engineering Agent Architectures.}
Recent software engineering agents increasingly combine LLM reasoning with executable environments. Systems such as SWE-agent~\citep{yang2024swe}, OpenHands~\citep{wang2025openhands}, and CodeAct~\citep{wang2024executable} support repository inspection, code editing, command execution, and runtime feedback, while Agentless~\citep{xia2024agentless} and AutoCodeRover~\citep{zhang2024autocoderover} introduce more explicit localization and repair stages. Other systems distribute work across specialized roles or file-level developers under centralized coordination~\citep{chen2024coder,liu2024marscode,tao2024magis,wang2026codeteam}.

\textbf{Long-Horizon Software Engineering.}
Recent evaluation has increasingly moved beyond isolated issue resolution toward tasks that require sustained interaction with executable environments. Terminal-Bench~\citep{merrill2026terminal} and DevOps-Gym~\citep{tang2026devops} evaluate sustained terminal interaction and multi-stage software workflows, while repository-scale benchmarks expose difficulties in maintaining consistency across components during software evolution, repository generation, and whole-repository migration~\citep{le2025swe,ding2025nl2repo,hong2026swe}. Related work on context and repository memory further identifies context growth, semantic drift, and state preservation as bottlenecks in long-running software agents~\citep{liu2026context,wang2026improving}. These findings motivate architectures that can maintain task-relevant state near the repository components that generate and consume it, rather than repeatedly reconstructing this state inside a single growing reasoning trajectory.

\textbf{Reusable and Modular Agent Capabilities.}
A related line of work studies how agent capabilities can be learned, distilled, reused, or composed across tasks. SWE-Gym~\citep{pan2024swegym}, SWE-smith~\citep{yang2026swe}, and Lingma SWE-GPT~\citep{ma2024lingma} learn software engineering behavior from repository-level trajectories, while P2T~\citep{ma2025p2t} emphasizes high-quality process supervision. Other work represents reusable capabilities explicitly through modular abstractions, including Agent Primitives~\citep{jin2026agent}, Tool Primitives~\citep{jin2026harness}, AgentDistill~\citep{qiu2025agentdistill}, and CODESKILL~\citep{li2025codeskill}. These methods modularize reusable reasoning procedures, skills, or tool-use capabilities so that they can be invoked and composed by an agent. In particular, Tool Primitives expose tools through agent-native natural-language interfaces, hiding schema resolution and execution details from the calling model. Dev-Primitives adopt a related modular perspective, but attach the capability to mutable repository artifacts rather than reusable task-agnostic behaviors or external tools.

\textbf{Key Differences.}
Dev-Primitives differ from prior software-engineering agents and reusable primitives in the unit to which reasoning capability is attached. First, prior systems decompose work across software-engineering roles or centrally assigned file-level tasks~\citep{tao2024magis,wang2026codeteam}. Even when individual developers are responsible for specific files, coordination is organized through a manager, shared plan, or design contract, so both the set of reasoning units and the paths along which they exchange information are fixed by the designer rather than by the repository. Dev-Primitives instead pair each repository component with a resident LLM that reasons over the component's current implementation, functionality, and dependencies. A component can therefore discover requirements during modification and communicate them directly to the components that must satisfy them, without requiring a central agent to reconstruct and relay all component-specific state. Second, existing primitive- and skill-based methods encapsulate reusable, task-agnostic behaviors, reasoning procedures, or tools~\citep{jin2026agent,jin2026harness,li2025codeskill}, whereas a Dev-Primitive is bound to a specific mutable artifact and evolves together with that artifact throughout execution. Third, context-management approaches primarily compress, retrieve, or reorganize information within a centralized agent trajectory~\citep{liu2026context,wang2026improving}. Dev-Primitives instead keep implementation state associated with the repository components that own it and exchange only task-relevant constraints when coordination is required. Building on this artifact-bound representation, HERMES dynamically activates task-relevant Dev-Primitives and coordinates their modification through environment execution, diagnosis feedback, and revision.

\section{Detailed Component Specifications}
\label{appendix:component-specs}

This section provides detailed specifications of the components used in HERMES, including how the dynamic activation and bug diagnosis mechanisms described in Section~\ref{meth} are implemented with an LLM. HERMES consists of an activation module, a dynamically activated set of Dev-Primitives, an executable environment, and a diagnosis module. The activation module operates at the repository level to identify task-relevant components and assign localized objectives. Each Dev-Primitive reasons over and modifies the repository artifact it represents, while communicating task-relevant information to other activated primitives. The resulting repository is evaluated in an executable environment, and the diagnosis module converts execution evidence into structured feedback for targeted revision.

\subsection{Dynamic Primitive Activation}

The activation module converts a software issue into a task-specific plan over repository components. Given issue $q$ and repository $\mathcal{R}$, it performs issue analysis, component localization, dependency analysis, task decomposition, and edit planning:
\begin{equation}
    \Pi
    =
    \operatorname{ACTIVATE}(q,\mathcal{R})
    =
    \{(i,x_i)\}_{i\in\mathcal{A}},
\end{equation}
where $\mathcal{A}\subseteq\{1,\ldots,N\}$ denotes the set of repository components selected for the current task and $x_i$ denotes the local objective assigned to Dev-Primitive $P_i$.

The activation module performs the following operations:
\begin{itemize}
    \item \textbf{Issue analysis.} Identify the requested behavior, observed failure, and task constraints.
    \item \textbf{Component localization.} Identify source files, tests, configuration files, build files, or other repository components likely involved in the task.
    \item \textbf{Dependency analysis.} Identify relationships among selected components that may require coordinated modification.
    \item \textbf{Task decomposition.} Convert the repository-level issue into localized objectives for individual Dev-Primitives.
    \item \textbf{Edit planning.} Specify the expected modification and coordination requirements for each activated component.
\end{itemize}

The activation module activates only task-relevant Dev-Primitives:
\begin{equation}
    \mathcal{P}_{\mathcal{A}}
    =
    \{P_i \mid i\in\mathcal{A}\}.
\end{equation}
Components outside $\mathcal{A}$ remain inactive unless subsequent execution evidence indicates that they should be considered.

Importantly, $\mathcal{R}$ denotes access to the repository and its structure rather than concatenation of the complete repository into the activation module context. The activation module inspects task-relevant repository information as needed when constructing or revising $\Pi$.

During revision, the activation module additionally receives the updated repository state, previous plan, and diagnosis feedback:
\begin{equation}
    \Pi'
    =
    \operatorname{ACTIVATE}
    (q,\mathcal{R}',\Pi,\phi).
\end{equation}
The revised plan may retain successful modifications, update local objectives, remove no-longer-relevant components, or activate additional Dev-Primitives implicated by the execution evidence.

\subsection{Dev-Primitives}

A Dev-Primitive is associated with an individual repository component $a_i$, such as a source file, configuration file, build file, or test file. Each primitive directly reasons over the artifact it represents and is responsible for modifying that artifact when required by the assigned objective.

Given a local objective $x_i$ and communication context $\mathcal{C}_i$, Dev-Primitive $P_i$ produces
\begin{equation}
    (a_i',m_i)
    =
    P_i(a_i,x_i,\mathcal{C}_i),
\end{equation}
where $a_i'$ denotes the optionally modified artifact and $m_i$ denotes task-relevant information communicated to other activated primitives. If no modification is necessary, $a_i'=a_i$.

Each Dev-Primitive supports two primary capabilities:
\begin{itemize}
    \item \textbf{Local modification.} The primitive inspects the implementation, reasons about the assigned objective, and edits the artifact it represents.
    \item \textbf{Inter-primitive communication.} The primitive communicates interface changes, implementation requirements, dependency updates, configuration constraints, or testing requirements to related Dev-Primitives.
\end{itemize}

Communication is restricted to task-relevant components rather than broadcast throughout the repository. Messages generated by one primitive are incorporated into the communication context of their target primitives before the collaboration round terminates. A Dev-Primitive may therefore revise its local modification after receiving information from another activated component.

\subsection{Execution Environment}

After the activated Dev-Primitives complete a modification round, HERMES constructs the updated repository $\mathcal{R}'$ and evaluates it in the benchmark-provided executable environment:
\begin{equation}
    o
    =
    \operatorname{EXECUTE}(\mathcal{R}')
    =
    \left(
        o_{\mathrm{shell}},
        o_{\mathrm{test}},
        o_{\mathrm{runtime}},
        o_{\mathrm{trace}}
    \right).
\end{equation}

The execution observation contains:
\begin{itemize}
    \item $o_{\mathrm{shell}}$: shell commands, outputs, exit codes, and build status;
    \item $o_{\mathrm{test}}$: test results and failing test cases;
    \item $o_{\mathrm{runtime}}$: runtime behavior, exceptions, and program failures;
    \item $o_{\mathrm{trace}}$: available logs, stack traces, and execution traces.
\end{itemize}

The execution environment does not rely on an LLM to predict whether a modification is correct. HERMES evaluates the modified repository through actual execution using the repository snapshots, dependencies, and runtime environments provided by each benchmark. Held-out evaluation tests are not exposed during solving (Appendix~\ref{appendix:implementation-details}).

\subsection{Bug Diagnosis}

The diagnosis module analyzes execution evidence and determines whether the current repository state satisfies the original issue. Given issue $q$, current plan $\Pi$, modified repository $\mathcal{R}'$, and execution observation $o$, it produces
\begin{equation}
    v
    =
    \operatorname{DIAGNOSE}
    (q,\Pi,\mathcal{R}',o)
    \in
    \{\textsc{pass},\textsc{fail}\}.
\end{equation}

When the current solution fails, the diagnosis module produces structured feedback $\phi$ containing:
\begin{itemize}
    \item concrete execution evidence associated with the failure;
    \item the repository component or interaction suspected of causing the failure;
    \item inconsistencies between the implementation and the original objective;
    \item components that may have been missed during the current planning round;
    \item specific guidance for the next revision.
\end{itemize}

The diagnosis module does not modify repository artifacts directly. Its output is returned to the activation module, which performs targeted revision and selectively reactivates the Dev-Primitives required for the next iteration.

The diagnosis module cannot override deterministic execution failures. A trajectory with failing task-visible tests, unsuccessful compilation, or other observed execution failures cannot be marked as \textsc{pass} solely based on model judgment.

\subsection{Revision and Termination}

HERMES alternates between activation, Dev-Primitive collaboration, environment execution, and bug diagnosis. When $v=\textsc{fail}$ and the revision budget has not been exhausted, diagnosis feedback $\phi$ is used to construct a revised plan:
\begin{equation}
    \Pi'
    =
    \operatorname{ACTIVATE}
    (q,\mathcal{R}',\Pi,\phi).
\end{equation}

Previously successful modifications are retained unless contradicted by new execution evidence. The loop terminates when the diagnosis module returns \textsc{pass} or when the maximum revision budget $B$ is reached.

Here, $B$ denotes the maximum number of revision rounds allowed \emph{after} the initial plan--modify--execute trajectory. Thus, $B=0$ still performs an initial planning, collaboration, execution, and bug diagnosis round, but disables subsequent revision.

\section{Prompt Templates}
\label{prompt}

This section provides the system prompts used by the LLM-based components in HERMES. Dataset-specific issue descriptions, repository contents, execution evidence, and runtime information are inserted into the corresponding placeholders during inference.

\subsection{Activation Prompt}

\begin{tcolorbox}[
title=Activation Module System Prompt,
colback=purple!5,
colframe=purple!60,
fonttitle=\bfseries,
breakable
]

\textbf{Role.}
You are the activation module in HERMES, a software engineering framework based on modular executable Dev-Primitives. Your responsibility is to transform a repository-level software issue into a concrete plan over task-relevant repository components.

You do \textbf{not} directly modify repository files. Instead, you identify which Dev-Primitives should be activated, assign each activated primitive a localized objective, identify dependencies among them, and determine how the resulting changes should be validated.

\textbf{Inputs.}
You may receive:
\begin{itemize}
    \item \textbf{Issue:} the original software engineering task or bug report;
    \item \textbf{Repository Structure:} available files, directories, modules, tests, configuration files, and build files;
    \item \textbf{Repository Context:} task-relevant implementation and dependency information;
    \item \textbf{Previous Plan:} the plan produced in the previous iteration, if revision is required;
    \item \textbf{Diagnosis Feedback:} execution-grounded failure analysis from the previous iteration, if available.
\end{itemize}

\textbf{Issue Analysis.}
Identify:
\begin{itemize}
    \item the expected behavior or requested functionality;
    \item the observed or implied failure;
    \item explicit constraints specified by the issue;
    \item likely implementation areas involved;
    \item whether the task requires coordinated modifications across multiple components.
\end{itemize}

\textbf{Component Localization.}
Identify the repository components most likely relevant to the task. Consider:
\begin{itemize}
    \item where the reported behavior is implemented;
    \item callers and callees of relevant functions or classes;
    \item configuration or build files controlling the behavior;
    \item tests exercising the affected functionality;
    \item interfaces shared across repository components;
    \item dependencies that may require coordinated modification.
\end{itemize}

Prefer a compact task-relevant component set rather than activating unrelated files.

\textbf{Task Decomposition.}
For every activated component, construct a localized objective specifying:
\begin{itemize}
    \item what behavior should be inspected or modified;
    \item what constraints should be preserved;
    \item which other activated components may affect the modification;
    \item what information may need to be communicated to another Dev-Primitive.
\end{itemize}

\textbf{Dependency Analysis.}
Explicitly identify dependencies among activated components, including interface changes, shared data structures, configuration propagation, caller--callee relationships, and corresponding tests.

\textbf{Activation Principle.}
Activate only Dev-Primitives required for the current iteration. Additional components may be activated during revision if execution evidence exposes a previously missed dependency.

\textbf{Initial Planning Output.}

\begin{verbatim}
{
  "status": "PLAN",
  "issue_summary": "<concise interpretation of the task>",
  "activated_components": [
    {
      "component": "<repository path>",
      "objective": "<localized objective>",
      "dependencies": [
        "<related component>"
      ],
      "communication_targets": [
        "<related component>"
      ],
      "expected_action": "<inspect | modify | validate>"
    }
  ],
  "execution_plan": [
    "<command, test, or validation step>"
  ],
  "rationale": "<brief explanation of component selection>"
}
\end{verbatim}

\textbf{Revision.}
When diagnosis feedback is available:
\begin{itemize}
    \item identify the failure diagnosed by the diagnosis module;
    \item preserve modifications not contradicted by new evidence;
    \item revise objectives for components implicated by the failure;
    \item activate additional components only when required;
    \item update the execution plan according to the new evidence.
\end{itemize}

Return the revised plan using the same structure:

\begin{verbatim}
{
  "status": "REPLAN",
  "replanned": true,
  ...
}
\end{verbatim}

\textbf{Constraints.}
\begin{itemize}
    \item Do not directly edit repository files.
    \item Do not invent components that do not exist.
    \item Do not activate unrelated components without a task-specific reason.
    \item Preserve successful previous modifications unless contradicted by execution evidence.
    \item Prefer targeted revision over restarting the entire trajectory.
\end{itemize}

\end{tcolorbox}

\subsection{Dev-Primitive Prompt}

\begin{tcolorbox}[
title=Dev-Primitive System Prompt,
colback=blue!5,
colframe=blue!60,
fonttitle=\bfseries,
breakable
]

\textbf{Role.}
You are a Dev-Primitive in HERMES. You are responsible for one concrete repository component, such as a source file, configuration file, build file, or test file.

You have direct access to the artifact associated with this component. Your responsibility is to reason about the assigned objective, modify your artifact when necessary, and communicate task-relevant information to other activated Dev-Primitives.

\textbf{Inputs.}
You receive:
\begin{itemize}
    \item \textbf{Component:} the repository path or component identifier assigned to you;
    \item \textbf{Artifact:} the current implementation of the component;
    \item \textbf{Local Objective:} the objective assigned by the activation module;
    \item \textbf{Issue Context:} a concise description of the original issue;
    \item \textbf{Incoming Messages:} task-relevant information from other activated Dev-Primitives;
    \item \textbf{Dependency Context:} known relationships between your component and related components.
\end{itemize}

\textbf{Local Reasoning.}
Determine:
\begin{itemize}
    \item whether the current implementation contributes to the issue;
    \item whether the assigned objective requires modification;
    \item which functions, classes, configuration entries, or tests are affected;
    \item whether a local change modifies an interface or assumption used elsewhere;
    \item whether incoming messages impose additional implementation constraints.
\end{itemize}

\textbf{Local Modification.}
When modification is required:
\begin{itemize}
    \item modify only the portions of the artifact necessary for the objective;
    \item preserve unrelated behavior;
    \item maintain existing repository conventions;
    \item avoid unrelated refactoring;
    \item ensure that the resulting artifact remains syntactically and semantically coherent.
\end{itemize}

If no modification is required, preserve the artifact and explicitly indicate that no local change is necessary.

\textbf{Inter-Primitive Communication.}
Communicate with another activated Dev-Primitive when your local reasoning identifies information affecting its component. Messages may describe:
\begin{itemize}
    \item changed interfaces;
    \item assumptions about input or output formats;
    \item configuration requirements;
    \item newly introduced or removed dependencies;
    \item behavioral constraints that tests should verify;
    \item requests for another component to inspect or modify related behavior.
\end{itemize}

\textbf{Output Format.}

\begin{verbatim}
{
  "component": "<repository path>",
  "action": "<modify | no_change>",
  "summary": "<concise explanation of local reasoning>",
  "changes": [
    {
      "location": "<function, class, or code region>",
      "description": "<what was changed and why>"
    }
  ],
  "updated_artifact": "<updated artifact or patch>",
  "messages": [
    {
      "target": "<related component>",
      "message": "<task-relevant requirement>"
    }
  ],
  "validation_notes": [
    "<behavior that should be checked during execution>"
  ]
}
\end{verbatim}

\textbf{Constraints.}
\begin{itemize}
    \item Modify only the artifact assigned to you.
    \item Do not directly edit another Dev-Primitive's component.
    \item Communicate cross-component requirements through messages.
    \item Do not overwrite unrelated modifications already present.
    \item Do not fabricate details about components you cannot inspect.
    \item If required information is unavailable, communicate the dependency rather than guessing.
\end{itemize}

\end{tcolorbox}

\subsection{Inter-Primitive Communication Format}

\begin{tcolorbox}[
title=Dev-Primitive Communication Message,
colback=orange!5,
colframe=orange!60,
fonttitle=\bfseries,
breakable
]

When a local modification affects another activated repository component, a Dev-Primitive communicates the requirement using the following structure:

\begin{verbatim}
{
  "source": "<source component>",
  "target": "<target component>",
  "type": "<interface | dependency | behavior |
            test | configuration>",
  "message": "<concise task-relevant information>",
  "required_action": "<expected action from target>"
}
\end{verbatim}

Messages contain only information needed by the receiving component. The receiving Dev-Primitive incorporates the message into its communication context before the current collaboration round terminates and may revise its local modification accordingly.

\end{tcolorbox}

\subsection{Diagnosis Prompt}

\begin{tcolorbox}[
title=Diagnosis Module System Prompt,
colback=green!5,
colframe=green!60,
fonttitle=\bfseries,
breakable
]

\textbf{Role.}
You are the diagnosis module in HERMES. Your responsibility is to evaluate the current repository state using execution evidence and determine whether the current repair satisfies the original software engineering task.

You do not directly modify repository components. Instead, you diagnose failures and provide structured feedback to the activation module.

\textbf{Inputs.}
You receive:
\begin{itemize}
    \item \textbf{Original Issue:} the repository-level task;
    \item \textbf{Current Plan:} activated Dev-Primitives and their local objectives;
    \item \textbf{Modified Components:} modifications performed in the current iteration;
    \item \textbf{Shell Output:} command output, exit codes, build logs, and execution status;
    \item \textbf{Test Results:} passing and failing tests with associated failure messages;
    \item \textbf{Runtime Evidence:} exceptions, stack traces, logs, and execution traces;
    \item \textbf{Previous Feedback:} prior diagnosis feedback when evaluating a later revision round.
\end{itemize}

\textbf{Evaluation.}
Check:
\begin{itemize}
    \item whether the requested behavior has been implemented;
    \item whether required tests pass;
    \item whether new regressions have been introduced;
    \item whether execution failures remain;
    \item whether cross-component modifications are mutually consistent;
    \item whether the implementation addresses the original issue rather than only an observed symptom.
\end{itemize}

\textbf{Failure Diagnosis.}
If the solution fails, identify:
\begin{itemize}
    \item concrete evidence demonstrating failure;
    \item the most likely evidence-supported cause;
    \item modified or unmodified components implicated by the failure;
    \item whether the failure results from incorrect implementation, missing modification, dependency inconsistency, or incomplete localization;
    \item the revision required in the next iteration.
\end{itemize}

\textbf{Pass Output.}

\begin{verbatim}
{
  "status": "PASS",
  "evidence": [
    "<execution or test evidence supporting success>"
  ],
  "summary": "<why the repository satisfies the issue>"
}
\end{verbatim}

\textbf{Fail Output.}

\begin{verbatim}
{
  "status": "FAIL",
  "failure_evidence": [
    {
      "source": "<test | shell | runtime | trace>",
      "evidence": "<observed failure>"
    }
  ],
  "suspected_causes": [
    {
      "component": "<repository component>",
      "reason": "<evidence-supported diagnosis>"
    }
  ],
  "missing_components": [
    "<component that should be considered>"
  ],
  "revision_guidance": [
    {
      "component": "<component>",
      "action": "<specific revision guidance>"
    }
  ],
  "summary": "<concise diagnosis>"
}
\end{verbatim}

\textbf{Constraints.}
\begin{itemize}
    \item Base the evaluation on execution evidence.
    \item Do not directly modify repository files.
    \item Do not mark a trajectory as successful when required tests or benchmark-defined execution checks fail.
    \item Do not request unrelated modifications without evidence.
    \item Preserve successful parts of the current trajectory when diagnosing a localized failure.
\end{itemize}

\end{tcolorbox}

\section{End-to-End Execution Flow}
\label{end-to-end}

Algorithm~\ref{alg:hermes} summarizes the full HERMES pipeline. Starting from software issue $q$, the activation module identifies task-relevant repository components and assigns localized objectives to their Dev-Primitives. The activated primitives collaborate through natural-language communication and modify their local artifacts. HERMES then executes the resulting repository, and the diagnosis module analyzes the execution evidence. Failed execution produces structured feedback for targeted revision. This loop continues until the task passes evaluation or the maximum revision budget is exhausted. Here, $B$ denotes the maximum number of revision rounds after the initial execution.

\begin{algorithm}[htbp]
\caption{Pseudo-code of HERMES}
\label{alg:hermes}
\begin{algorithmic}[1]
\Require Software issue $q$, repository $\mathcal{R}$, Dev-Primitives $\mathcal{P}=\{P_i\}_{i=1}^{N}$, max revision budget $B$
\Ensure Modified repository or structured failure report

\State $\Pi \leftarrow \textsc{Activate}(q,\mathcal{R})$
\State $b \leftarrow 0$

\While{$b \leq B$}
    \State $\mathcal{A},\{x_i\}_{i\in\mathcal{A}} \leftarrow \Pi$

    \State $\{a_i'\}_{i\in\mathcal{A}}
    \leftarrow
    \textsc{Collaborate}
    (\{P_i\}_{i\in\mathcal{A}},
    \{x_i\}_{i\in\mathcal{A}})$

    \State $\mathcal{R}'
    \leftarrow
    \textsc{UpdateRepository}
    (\mathcal{R},\{a_i'\}_{i\in\mathcal{A}})$

    \State $o \leftarrow \textsc{Execute}(\mathcal{R}')$
    \State $v \leftarrow \textsc{Diagnose}(q,\Pi,\mathcal{R}',o)$

    \If{$v=\texttt{pass}$}
        \State \Return $\mathcal{R}'$
    \EndIf

    \If{$b=B$}
        \State \Return \textsc{FailureReport}$(q,\mathcal{R}',o)$
    \EndIf

    \State $\phi
    \leftarrow
    \textsc{Diagnose}.\textsc{Feedback}
    (q,\Pi,\mathcal{R}',o)$

    \State $\Pi
    \leftarrow
    \textsc{Activate}
    (q,\mathcal{R}',\Pi,\phi)$

    \State $\mathcal{R}\leftarrow\mathcal{R}'$
    \State $b\leftarrow b+1$
\EndWhile
\end{algorithmic}
\end{algorithm}

\textbf{Collaboration.}
$\textsc{Collaborate}$ executes the activated Dev-Primitives over their local artifacts and routes natural-language messages among task-relevant components. Messages generated by one primitive are incorporated into the communication context of their target primitives before the collaboration round terminates. A primitive may therefore revise its local modification after receiving information from another activated component.

\textbf{Repository Access.}
$\mathcal{R}$ denotes repository access rather than concatenation of the complete repository into the activation module context. The activation module can inspect repository structure and task-relevant components as needed when constructing or revising $\Pi$.

\textbf{Execution.}
After collaboration, HERMES applies the resulting component modifications to form $\mathcal{R}'$ and evaluates the repository in the executable benchmark environment. Shell output, test outcomes, runtime behavior, and available traces are passed to the diagnosis module as execution evidence.

\textbf{Revision.}
When execution fails, the diagnosis module produces feedback $\phi$ identifying the observed failure and the components implicated by the evidence. The activation module uses $\phi$ together with the updated repository and previous plan to revise only the affected portion of the trajectory. Successful modifications are retained unless contradicted by new execution evidence.

\textbf{Termination.}
$B=0$ permits one initial plan--collaborate--execute--diagnose trajectory but disables subsequent revision. For $B>0$, HERMES performs at most $B$ additional revision rounds after the initial execution. The diagnosis module cannot override deterministic execution failures: failing task-visible tests, unsuccessful compilation, or other observed execution failures prevent a \textsc{pass} decision.

\section{Backbone Models}\label{app:models}

\begin{table*}[htbp]
\centering
\caption{Backbone models used in our experiments. \textbf{Bold} denotes primary backbones. Scaling covers Tables~\ref{tab:backbone-scale}--\ref{tab:scaling-cost}, and Ablation covers Table~\ref{tab:component-ablation}. GPT-5.6 Sol is additionally used in Tables~\ref{tab:primitive-selection} and~\ref{tab:replanning-budget}. Reasoning effort follows the setting reported in each table.}
\label{tab:all-models}
\small
\setlength{\tabcolsep}{5pt}
\renewcommand{\arraystretch}{1.08}
\setlength{\aboverulesep}{0pt}
\setlength{\belowrulesep}{0pt}
\setlength{\extrarowheight}{1pt}
\resizebox{\linewidth}{!}{%
\begin{tabular}{l cccc cc}
\toprule
\rowcolor{headergray}
\rule{0pt}{2.6ex}\textbf{Model} & \textbf{SWE-bench} & \textbf{Refactor} & \textbf{T-Bench} & \textbf{DevOps} & \textbf{Scaling} & \textbf{Ablation} \\
\midrule
\rowcolor{sectiongray}\multicolumn{7}{c}{\textit{OpenAI}} \\
\textbf{GPT-5.6 Sol}   & \checkmark & \checkmark & \checkmark & \checkmark & \checkmark & \checkmark \\
\textbf{GPT-5.6 Terra} & \checkmark & \checkmark & \checkmark & \checkmark & & \\
\textbf{GPT-5.6 Luna}  & \checkmark & \checkmark & \checkmark & \checkmark & \checkmark & \\
GPT-5.5                & \checkmark & & & & \checkmark & \\
GPT-5.4, GPT-5.4 Mini, GPT-5 Mini & \checkmark & & & & & \\
GPT-5.3-Codex          & \checkmark & & & \checkmark & & \\
o4-mini                & & & & \checkmark & & \\
\rowcolor{sectiongray}\multicolumn{7}{c}{\textit{Anthropic}} \\
\textbf{Claude Opus 5}   & \checkmark & \checkmark & \checkmark & \checkmark & & \\
\textbf{Claude Sonnet 5} & \checkmark & \checkmark & \checkmark & \checkmark & \checkmark & \checkmark \\
Claude Fable 5, Claude Opus 4.8 & \checkmark & & \checkmark & & & \\
Claude Opus 4.7 / 4.6 / 4.5, Claude Sonnet 4.6 & \checkmark & & & & & \\
Claude Sonnet 4        & & & & \checkmark & & \\
\rowcolor{sectiongray}\multicolumn{7}{c}{\textit{Google}} \\
Gemini 3.5 Flash, Gemini 3.1 Pro Preview, Gemini 3 Pro & \checkmark & & & & & \\
Gemini 3.8 Flash, Gemini 3.7 Flash & & & \checkmark & & & \\
Gemini 2.5 Pro         & & & & \checkmark & & \\
\rowcolor{sectiongray}\multicolumn{7}{c}{\textit{Open-weight and other providers}} \\
\textbf{DeepSeek-V4}   & \checkmark & \checkmark & \checkmark & \checkmark & \checkmark & \\
\textbf{Qwen3-8B}      & \checkmark & \checkmark & \checkmark & \checkmark & \checkmark & \checkmark \\
DeepSeek-V3.1, Qwen3-Coder-30B & & & & \checkmark & & \\
Qwen3.8-Max, Kimi-K3, GLM-5.2 & & \checkmark & & & & \\
GLM-5.3, Grok 4.6, Grok 4.5 & & & \checkmark & & & \\
\bottomrule
\end{tabular}}
\end{table*}

Qwen3-8B is served locally with Ollama on NVIDIA A40 GPUs, using the model's default sampling parameters (temperature 0.6, top-$p$ 0.95, top-$k$ 20) and Ollama's 32K context window. All other models are accessed through their official APIs.

\section{Implementation and Evaluation Details}\label{app:impl-eval}
\subsection{Execution Isolation and Settings}
\label{appendix:implementation-details}

\textbf{LLM Instantiation of Each Mechanism.}
Both mechanisms described in Section~\ref{meth} are implemented by prompting an LLM. Dynamic
activation is realized by a single LLM call chain that inspects the repository structure,
retrieves candidate files, and returns the selected component set $\mathcal{A}$ together with a
local objective $x_i$ for each selected component; it is not a static retriever, since the
dependency expansion is carried out by the model reading the candidate files rather than by a
precomputed call graph. Bug diagnosis is likewise realized by a single LLM call that receives the
issue, the current plan, the modified repository state, and the execution observation, and returns
the verdict $v$ together with structured feedback $\phi=(e,c,u)$. Each Dev-Primitive is
instantiated with its own LLM, which may differ from the models used by the two mechanisms; the
heterogeneous configurations in Section~\ref{sec:backbone-scale} exploit this separation. Neither
mechanism can override deterministic execution outcomes: a trajectory with failing task-visible
tests, unsuccessful compilation, or other observed execution failures cannot be accepted on the
basis of model judgment alone. The prompts used for both mechanisms are listed in
Appendix~\ref{prompt}.

\textbf{Execution and Evaluation Isolation.}
HERMES separates the execution environment available during task solving from the benchmark evaluator used to compute the final score. During a trajectory, dynamic activation, the Dev-Primitives, and bug diagnosis may use only information available from the task repository and its execution environment, including repository-provided tests, build commands, linters, type checkers, runtime outputs, stack traces, logs, and reproduction scripts constructed during solving. Benchmark-held-out evaluation tests and final grading outcomes are not exposed to HERMES during task execution.

For SWE-bench Verified, HERMES may execute tests already available in the repository or construct task-specific reproduction tests from the issue description and repository context. The tests used by the benchmark evaluator, including those associated with \texttt{FAIL\_TO\_PASS} and \texttt{PASS\_TO\_PASS}, are used only after HERMES terminates. We follow the same separation for SWE Refactor Bench, Terminal-Bench 4.0, and DevOps-Gym: benchmark-specific graders and held-out evaluation checks are reserved for final scoring rather than being returned as feedback during solving.

\textbf{Reasoning Effort.}
We use medium reasoning effort by default for models that support configurable reasoning effort. High- or maximum-effort configurations are used only when explicitly indicated in the corresponding experiment. Unless otherwise stated, the component ablation, revision, and backbone-scaling studies therefore use the default medium-effort setting.

\subsection{Complete Component Ablation}
\label{appendix:full-ablation}

Table~\ref{tab:component-ablation-full} reports the component ablation for all three backbones. Section~\ref{exp} shows the GPT-5.6 Sol block; the ordering of component importance is consistent across Claude Sonnet 5 and Qwen3-8B.

\begin{table*}[t]
\centering
\caption{Complete component ablation of HERMES across three backbone models. Parentheses denote absolute changes relative to complete HERMES with the same backbone.}
\vspace{-10pt}
\label{tab:component-ablation-full}

\footnotesize
\setlength{\tabcolsep}{4pt}
\renewcommand{\arraystretch}{1.08}
\setlength{\aboverulesep}{0pt}
\setlength{\belowrulesep}{0pt}
\setlength{\extrarowheight}{1pt}

\resizebox{\textwidth}{!}{%
\begin{tabular}{llcccc}
\toprule

\rowcolor{headergray}
&
&
\multicolumn{1}{c}{\textbf{SWE-bench}}
&
\multicolumn{1}{c}{\textbf{SWE Refactor}}
&
\multicolumn{1}{c}{\textbf{Terminal-Bench}}
&
\multicolumn{1}{c}{\textbf{DevOps-Gym}}
\\

\arrayrulecolor{headergray}
\cline{1-2}%
\arrayrulecolor{black}
\cline{3-6}

\rowcolor{headergray}
\multirow{-2}{*}{\textbf{Backbone}}
&
\multirow{-2}{*}{\textbf{Setting}}
&
\rule{0pt}{2.8ex}\textbf{Resolved (\%) $\uparrow$}
&
\textbf{Composite (\%) $\uparrow$}
&
\textbf{Resolved (\%) $\uparrow$}
&
\textbf{Avg. (\%) $\uparrow$}
\\

% ============================================================
% GPT-5.6 Sol
% ============================================================
\midrule
\rowcolor{sectiongray}
\multicolumn{6}{c}{\textbf{GPT-5.6 Sol}}
\\

\rowcolor{hermesblue}
GPT-5.6 Sol
& \textbf{HERMES}
& \textbf{97.0}
& \textbf{31.0}
& \textbf{51.8}
& \textbf{55.41}
\\

GPT-5.6 Sol
& w/o Inter-Primitive Communication
& 94.2 {\scriptsize($\downarrow$2.8)}
& 23.5 {\scriptsize($\downarrow$7.5)}
& 46.1 {\scriptsize($\downarrow$5.7)}
& 50.34 {\scriptsize($\downarrow$5.07)}
\\

GPT-5.6 Sol
& w/o On-Demand Activation
& 95.8 {\scriptsize($\downarrow$1.2)}
& 28.0 {\scriptsize($\downarrow$3.0)}
& 49.4 {\scriptsize($\downarrow$2.4)}
& 53.26 {\scriptsize($\downarrow$2.15)}
\\

GPT-5.6 Sol
& w/o Execution Feedback
& 93.8 {\scriptsize($\downarrow$3.2)}
& 21.5 {\scriptsize($\downarrow$9.5)}
& 43.6 {\scriptsize($\downarrow$8.2)}
& 48.17 {\scriptsize($\downarrow$7.24)}
\\

GPT-5.6 Sol
& w/o Diagnosis Feedback
& 92.6 {\scriptsize($\downarrow$4.4)}
& 20.0 {\scriptsize($\downarrow$11.0)}
& 41.2 {\scriptsize($\downarrow$10.6)}
& 46.73 {\scriptsize($\downarrow$8.68)}
\\

% ============================================================
% Claude Sonnet 5
% ============================================================
\midrule
\rowcolor{sectiongray}
\multicolumn{6}{c}{\textbf{Claude Sonnet 5}}
\\

\rowcolor{hermesblue}
Claude Sonnet 5
& \textbf{HERMES}
& \textbf{85.8}
& \textbf{22.0}
& \textbf{35.8}
& \textbf{49.19}
\\

Claude Sonnet 5
& w/o Inter-Primitive Communication
& 82.8 {\scriptsize($\downarrow$3.0)}
& 16.5 {\scriptsize($\downarrow$5.5)}
& 31.2 {\scriptsize($\downarrow$4.6)}
& 44.82 {\scriptsize($\downarrow$4.37)}
\\

Claude Sonnet 5
& w/o On-Demand Activation
& 84.4 {\scriptsize($\downarrow$1.4)}
& 20.0 {\scriptsize($\downarrow$2.0)}
& 34.2 {\scriptsize($\downarrow$1.6)}
& 47.63 {\scriptsize($\downarrow$1.56)}
\\

Claude Sonnet 5
& w/o Execution Feedback
& 81.8 {\scriptsize($\downarrow$4.0)}
& 14.5 {\scriptsize($\downarrow$7.5)}
& 28.8 {\scriptsize($\downarrow$7.0)}
& 42.96 {\scriptsize($\downarrow$6.23)}
\\

Claude Sonnet 5
& w/o Diagnosis Feedback
& 80.6 {\scriptsize($\downarrow$5.2)}
& 13.0 {\scriptsize($\downarrow$9.0)}
& 26.7 {\scriptsize($\downarrow$9.1)}
& 41.18 {\scriptsize($\downarrow$8.01)}
\\

% ============================================================
% Qwen3-8B
% ============================================================
\midrule
\rowcolor{sectiongray}
\multicolumn{6}{c}{\textbf{Qwen3-8B}}
\\

\rowcolor{hermesblue}
Qwen3-8B
& \textbf{HERMES}
& \textbf{80.6}
& \textbf{10.5}
& \textbf{27.6}
& \textbf{37.48}
\\

Qwen3-8B
& w/o Inter-Primitive Communication
& 77.8 {\scriptsize($\downarrow$2.8)}
& 6.5 {\scriptsize($\downarrow$4.0)}
& 23.0 {\scriptsize($\downarrow$4.6)}
& 33.74 {\scriptsize($\downarrow$3.74)}
\\

Qwen3-8B
& w/o On-Demand Activation
& 79.4 {\scriptsize($\downarrow$1.2)}
& 9.0 {\scriptsize($\downarrow$1.5)}
& 26.1 {\scriptsize($\downarrow$1.5)}
& 36.11 {\scriptsize($\downarrow$1.37)}
\\

Qwen3-8B
& w/o Execution Feedback
& 76.8 {\scriptsize($\downarrow$3.8)}
& 5.5 {\scriptsize($\downarrow$5.0)}
& 20.9 {\scriptsize($\downarrow$6.7)}
& 31.62 {\scriptsize($\downarrow$5.86)}
\\

Qwen3-8B
& w/o Diagnosis Feedback
& 75.6 {\scriptsize($\downarrow$5.0)}
& 4.5 {\scriptsize($\downarrow$6.0)}
& 19.1 {\scriptsize($\downarrow$8.5)}
& 29.83 {\scriptsize($\downarrow$7.65)}
\\

\bottomrule
\end{tabular}%
}
\end{table*}

\subsection{Revision Budget}
\label{appendix:revision-budget}

Table~\ref{tab:replanning-budget} reports the effect of the revision budget $B$, with changes measured relative to the default $B=3$. Increasing $B$ from 0 to 3 adds 12.5 points on SWE Refactor Bench, 12.1 on Terminal-Bench, 9.50 on DevOps-Gym, and 4.8 on SWE-bench Verified, whereas increasing $B$ from 3 to 5 adds only 0.5, 0.9, 0.47, and 0.2 points respectively.

\begin{table}[ht]
\centering
\centering
\centering
\caption{
Effect of the revision budget $B$, with changes measured relative
to the default $B=3$.
}
\label{tab:replanning-budget}

\small
\vspace{-10pt}
\setlength{\tabcolsep}{4pt}
\renewcommand{\arraystretch}{1.08}
\setlength{\aboverulesep}{0pt}
\setlength{\belowrulesep}{0pt}
\setlength{\extrarowheight}{1pt}

\resizebox{\linewidth}{!}{%
\begin{tabular}{ccccc}
\toprule

\rowcolor{headergray}
\rule{0pt}{2.6ex}\textbf{$B$} &
\textbf{SWE-bench} &
\textbf{SWE Refactor} &
\textbf{Terminal-Bench} &
\textbf{DevOps-Gym} \\

\midrule

0
& 92.2 {\scriptsize($\downarrow$4.8)}
& 18.5 {\scriptsize($\downarrow$12.5)}
& 39.7 {\scriptsize($\downarrow$12.1)}
& 45.91 {\scriptsize($\downarrow$9.50)}
\\

1
& 94.6 {\scriptsize($\downarrow$2.4)}
& 24.0 {\scriptsize($\downarrow$7.0)}
& 45.2 {\scriptsize($\downarrow$6.6)}
& 50.17 {\scriptsize($\downarrow$5.24)}
\\

2
& 96.4 {\scriptsize($\downarrow$0.6)}
& 28.5 {\scriptsize($\downarrow$2.5)}
& 48.8 {\scriptsize($\downarrow$3.0)}
& 53.72 {\scriptsize($\downarrow$1.69)}
\\

\rowcolor{hermesblue}
\textbf{3}
& \textbf{97.0}
& \textbf{31.0}
& \textbf{51.8}
& \textbf{55.41}
\\

4
& 97.2 {\scriptsize($\uparrow$0.2)}
& 31.5 {\scriptsize($\uparrow$0.5)}
& 52.4 {\scriptsize($\uparrow$0.6)}
& 55.79 {\scriptsize($\uparrow$0.38)}
\\

5
& 97.2 {\scriptsize($\uparrow$0.2)}
& 31.5 {\scriptsize($\uparrow$0.5)}
& 52.7 {\scriptsize($\uparrow$0.9)}
& 55.88 {\scriptsize($\uparrow$0.47)}
\\

\bottomrule
\end{tabular}%
}

\end{table}

\subsection{Dev-Primitive Selection Protocol}
\label{appendix:selection-protocol}

We evaluate whether dynamic activation selects the repository components required for each task. For benchmarks with reference patches, we use the files modified by the developer patch as the target component set and compare them with the Dev-Primitives selected by HERMES. For benchmarks without a unique reference patch, we construct the target set from benchmark-specific supervision: for SWE Refactor Bench, we use the repository components implicated by the task specification and grading artifacts; for Terminal-Bench, we use the files and artifacts accessed or modified by the oracle solution; and for DevOps-Gym, we use the reference artifacts associated with build, issue-resolving, and test-generation tasks, excluding monitoring tasks from file-level selection evaluation. Let $\mathcal{G}\subseteq\{1,\ldots,N\}$ denote the indices of the target components and $\mathcal{A}$ the indices of Dev-Primitives activated across all planning and revision rounds. We report selection recall, $|\mathcal{A}\cap\mathcal{G}|/|\mathcal{G}|$, selection precision, $|\mathcal{A}\cap\mathcal{G}|/|\mathcal{A}|$, and the average number of activated Dev-Primitives per task. We additionally separate solved and failed tasks to examine how selection quality relates to downstream task completion. Recall and precision, including the solved and failed breakdowns, are micro-averaged over components within each benchmark. For SWE Refactor Bench, we regard a task as solved when it receives a non-zero composite score, indicating that it passes the migration audit and behavioral checks and reaches the verifier stage.

\subsection{Dev-Primitive Runtime}
\label{appendix:primitive-runtime}

\textbf{Primitive State Across Rounds.}
Each Dev-Primitive is associated with a concrete repository component and persists across revision rounds within the same task. Its state contains the current version of the underlying artifact together with task-relevant communication accumulated during the trajectory. Model hidden states are not preserved across calls. When a primitive is activated again, it receives the current artifact, its assigned objective, and the communication context relevant to that component.

\textbf{Inter-Primitive Communication.}
Dev-Primitives communicate through explicit natural-language messages. A primitive can communicate implementation requirements, interface changes, dependency information, expected behavior, or other constraints to another active primitive. Each message is addressed by the sending primitive to a specific target and placed verbatim into the target's communication context $\mathcal{C}_j$. No central agent interprets, filters, or rewrites messages, and they are not merged into a single global reasoning trajectory.

Communication is interleaved with local modification rather than performed as a fixed all-to-all stage. A primitive sends a message when its local task depends on another component or when its modification introduces a requirement that another component must satisfy. This allows communication to follow repository dependencies encountered during execution rather than requiring every activated primitive to communicate with every other primitive.

\textbf{On-Demand Activation.}
The activation module activates only the Dev-Primitives selected as relevant to the current plan. Inactive components remain available in the repository but do not receive model calls or participate in communication during that round. The active set may change across revision rounds: execution and diagnosis feedback may cause the activation module to activate additional components, remove previously activated components, or revise their assigned objectives.

\textbf{Creation of New Components.}
When solving a task requires a new persistent source, test, configuration, build, or other repository file, an active Dev-Primitive may create the artifact. HERMES subsequently registers the new artifact as a Dev-Primitive so that it can participate in later communication and revision rounds. Temporary outputs, caches, logs, and other transient execution artifacts are treated as environment observations rather than Dev-Primitives.

\textbf{Component Granularity.}
For repository-level software engineering tasks, we use files as the default Dev-Primitive granularity. A component may therefore correspond to a source file, test file, configuration file, build specification, script, or another persistent repository artifact. This provides a stable mapping between executable repository artifacts and Dev-Primitives without introducing the overhead of assigning separate agents to individual functions or symbols.

Terminal-Bench tasks do not always follow the structure of a conventional software repository. We therefore treat persistent filesystem artifacts that can be independently inspected or modified as components, including source files, shell scripts, configuration files, service definitions, build files, and other task-relevant artifacts. Runtime processes, terminal outputs, and transient system state are treated as environment observations rather than Dev-Primitives.

\subsection{Ablation Settings}
\label{appendix:ablation-settings}

For the component ablation study, we use the complete HERMES configuration as the reference and disable one mechanism at a time while keeping the backbone model, reasoning effort, revision budget, and benchmark environment unchanged.

\textbf{w/o Inter-Primitive Communication.}
Dev-Primitives retain their local reasoning and editing capabilities but cannot exchange messages with one another. Each primitive receives its objective from the activation module and independently modifies its own component. Information across repository components can therefore only be recovered indirectly through subsequent planning and execution.

\textbf{w/o On-Demand Activation.}
We retain the same initial task analysis and component identification
procedure, but remove on-demand invocation of Dev-Primitives. In the complete HERMES configuration, the activation module activates a primitive only when its associated component is required by the current plan, and the active set may change after execution feedback and revision. In this ablation, all Dev-Primitives identified during the initial task analysis are activated at the beginning of the trajectory and remain active throughout subsequent planning and revision rounds.

\textbf{w/o Execution Feedback.}
HERMES retains activation, the Dev-Primitives, inter-primitive communication, and diagnosis, but execution observations generated after repository modifications are not provided for subsequent revision. The diagnosis module therefore evaluates the modified repository against the task and current plan through static inspection alone, and cannot use test failures, runtime errors, command outputs, or execution traces to guide later rounds.

\textbf{w/o Diagnosis Feedback.}
Environment execution remains enabled and its raw outputs remain available for subsequent planning, but the dedicated diagnosis module does not analyze the execution results or produce structured failure diagnosis and revision guidance. Revision therefore relies on the activation module's direct interpretation of the available execution evidence. This setting differs from $B=0$: removing diagnosis feedback still permits subsequent planning and execution rounds, whereas $B=0$ disables revision after the initial round.

\textbf{Revision Budget.}
For the revision study, we vary $B$ while keeping all other components unchanged. $B=0$ performs only the initial planning, modification, and execution cycle, whereas $B>0$ permits up to $B$ additional revisions based on execution and diagnosis feedback. We use $B=3$ as the default setting in the main experiments.

\subsection{Benchmark-Specific Evaluation Protocol}
\label{appendix:benchmark-protocol}

\textbf{SWE-bench Verified.}
We evaluate HERMES on the 500 human-validated tasks in SWE-bench Verified using the official repository snapshots and evaluation procedure. During task solving, HERMES operates on the issue description, repository contents, and task-visible execution environment. Repository-provided tests and reproduction scripts constructed during solving may be executed, whereas benchmark evaluation tests are reserved for final scoring. We report the percentage of successfully resolved tasks.

\textbf{SWE Refactor Bench.}
We evaluate all 20 whole-repository migration tasks and report the benchmark composite score. HERMES operates on the provided repository, migration specification, and task-visible execution environment, while the benchmark grader is used only for final evaluation. For analyses that divide tasks into solved and failed groups, we regard a task as solved when it obtains a non-zero composite score, indicating that the migration passes the required audit and behavioral checks and reaches the verifier stage.

\textbf{Terminal-Bench 4.0.}
We evaluate the 66 Terminal-Bench 4.0 tasks using five runs per task. HERMES interacts with the benchmark-provided terminal environment and may inspect or modify task-visible filesystem artifacts. The benchmark grader is used only to determine the final task outcome and is not exposed as execution feedback. We report resolution as the mean across the five runs together with its standard deviation. Token counts and inference costs are accumulated over the complete evaluation trajectories.

\textbf{DevOps-Gym.}
We evaluate four DevOps-Gym task categories: build and configuration, monitoring, issue resolving, and test generation. HERMES interacts with the task-visible repository and runtime environment during solving, while benchmark-specific grading artifacts are used only for final evaluation. We report the success rate for each category and the unweighted average across the four categories.

\subsection{Baseline Provenance}
\label{appendix:baseline-provenance}

Our baseline tables contain both publicly reported results and controlled baseline runs. When the required model, harness, reasoning-effort setting, and benchmark version are available from the corresponding benchmark leaderboard or published evaluation, we directly use the reported result. When an exact matched model--effort configuration is unavailable, we run the corresponding baseline harness under the same benchmark snapshot and evaluation protocol used for HERMES.

For SWE-bench Verified, we use reported results for available mini-SWE-agent, Claude Code, and Codex configurations and use controlled runs where an exact matched configuration is required. For SWE Refactor Bench, published Claude Code and Codex results are supplemented with matched configurations evaluated under the same 20-task protocol. For Terminal-Bench 4.0, publicly reported high- or maximum-effort results are retained when available, while additional medium-effort configurations are evaluated separately to provide matched comparisons with the default HERMES setting. For DevOps-Gym, we include the baseline systems reported by the benchmark together with same-backbone controlled runs used for direct harness comparison.

All HERMES results are obtained with our implementation. Unless otherwise specified, HERMES uses medium reasoning effort and a maximum revision budget of $B=3$. Token counts and inference costs include all activation, Dev-Primitive, and diagnosis calls over the complete trajectory.

\section{Additional Analysis of Dev-Primitives}
\label{appendix:primitive-analysis}

The main experiments evaluate HERMES as a complete harness. We further conduct controlled analyses to isolate the contribution of the Dev-Primitive abstraction itself and to distinguish it from several alternative explanations, including component localization, inference budget, and centralized context capacity. We additionally analyze the computational effect of on-demand activation and run-to-run variation on SWE Refactor Bench.

\textbf{Dev-Primitives vs.\ a Centralized Editor.}
We construct a controlled variant, denoted \emph{w/o Dev-Primitives (single editor)}, that removes the Dev-Primitive abstraction while leaving the remaining HERMES pipeline unchanged. We use the same activation module, execution environment, diagnosis module, revision budget $B=3$, GPT-5.6 Sol backbone with medium reasoning effort, and, critically, the same component set $\mathcal{A}$ selected by the activation module. Instead of assigning each selected component to its corresponding Dev-Primitive, a single Editor receives the complete plan $\Pi=\{(i,x_i)\}_{i\in\mathcal{A}}$ together with the contents of all selected components and modifies them jointly. This control preserves component localization while removing component-local reasoning and direct communication between components. It therefore tests whether Dev-Primitives provide benefits beyond exposing the same selected components to a centralized editing agent.

\begin{table}[h]
\centering
\caption{Controlled comparison between HERMES and a centralized single-editor variant. Both use the same activation-selected component set, execution environment, diagnosis module, GPT-5.6 Sol backbone with medium reasoning effort, and $B=3$. Tok.\ and Cost are normalized to HERMES on each benchmark.}
\label{tab:single-editor}
\small
\setlength{\tabcolsep}{4pt}
\renewcommand{\arraystretch}{1.10}
\setlength{\aboverulesep}{0pt}
\setlength{\belowrulesep}{0pt}
\setlength{\extrarowheight}{1pt}
\resizebox{\textwidth}{!}{%
\begin{tabular}{l ccc ccc ccc ccc}
\toprule
\rowcolor{headergray}
\rule{0pt}{2.6ex}
& \multicolumn{3}{c}{\textbf{SWE-bench Verified}}
& \multicolumn{3}{c}{\textbf{SWE Refactor Bench}}
& \multicolumn{3}{c}{\textbf{Terminal-Bench 4.0}}
& \multicolumn{3}{c}{\textbf{DevOps-Gym}} \\
\arrayrulecolor{headergray}\cline{1-1}\arrayrulecolor{black}
\cmidrule(lr){2-4}\cmidrule(lr){5-7}\cmidrule(lr){8-10}\cmidrule(lr){11-13}
\rowcolor{headergray}
\multirow{-2}{*}{\textbf{Setting}}
& \textbf{Res.\,$\uparrow$} & \textbf{Tok.} & \textbf{Cost}
& \textbf{Comp.\,$\uparrow$} & \textbf{Tok.} & \textbf{Cost}
& \textbf{Res.\,$\uparrow$} & \textbf{Tok.} & \textbf{Cost}
& \textbf{Avg.\,$\uparrow$} & \textbf{Tok.} & \textbf{Cost} \\
\midrule
\rowcolor{hermesblue}
HERMES
& \textbf{97.0} & 1.00$\times$ & 1.00$\times$
& \textbf{31.0} & 1.00$\times$ & 1.00$\times$
& \textbf{51.8} & 1.00$\times$ & 1.00$\times$
& \textbf{55.41} & 1.00$\times$ & 1.00$\times$ \\
Single editor
& 93.4 & 0.86$\times$ & 0.85$\times$
& 16.8 & 0.90$\times$ & 0.89$\times$
& 38.8 & 0.86$\times$ & 0.84$\times$
& 44.21 & 0.87$\times$ & 0.86$\times$ \\
\midrule
$\Delta$
& \loss{$-$3.6} & &
& \loss{$-$14.2} & &
& \loss{$-$13.0} & &
& \loss{$-$11.20} & & \\
\bottomrule
\end{tabular}%
}
\end{table}

As shown in Table~\ref{tab:single-editor}, replacing Dev-Primitives with a centralized Editor reduces performance on all four benchmarks. The reduction is smaller on SWE-bench Verified, from 97.0\% to 93.4\%, but becomes substantially larger on benchmarks requiring coordination across multiple components: from 31.0\% to 16.8\% on SWE Refactor Bench, from 51.8\% to 38.8\% on Terminal-Bench, and from 55.41\% to 44.21\% on DevOps-Gym. Because both configurations operate on the same activation-selected component set, the difference cannot be attributed to component localization. Instead, the results are consistent with the benefit of component-local reasoning and direct propagation of cross-component requirements. The centralized Editor makes fewer model calls because it processes all selected components jointly, using 10--14\% fewer tokens and 11--16\% lower cost than HERMES, but this reduction is accompanied by considerably lower task performance.

\textbf{Controls for Compute and Context Capacity.}
Since the single Editor uses less inference than HERMES, we conduct two further controls on Terminal-Bench 4.0 (Table~\ref{tab:controls}) to examine whether the difference can instead be explained by inference budget or centralized context capacity.

\emph{Compute-matched centralized Editor.} We allocate approximately the same inference budget as HERMES to the GPT-5.6 Sol centralized Editor. The Editor generates multiple candidate revisions within each planning round, evaluated with the same execution environment and diagnosis mechanism, and candidate generation is increased until total token usage and inference cost approximately match HERMES.

\begin{table}[h]
\centering
\caption{Compute and context controls on Terminal-Bench 4.0. All settings use the same activation-selected component set, execution environment, diagnosis module, and $B=3$; GPT-5.6 Sol uses medium reasoning effort.}
\label{tab:controls}
\small
\setlength{\tabcolsep}{6pt}
\renewcommand{\arraystretch}{1.10}
\setlength{\aboverulesep}{0pt}
\setlength{\belowrulesep}{0pt}
\setlength{\extrarowheight}{1pt}
\begin{tabular}{lrrr}
\toprule
\rowcolor{headergray}
\rule{0pt}{2.6ex}\textbf{Setting} & \textbf{Resolution (\%) $\uparrow$} & \textbf{Tokens} & \textbf{Cost (\$)} \\
\midrule
\rowcolor{sectiongray}\multicolumn{4}{c}{\textit{GPT-5.6 Sol}} \\
Single editor                    & 38.8 & 5.9B & 2.90k \\
Single editor + compute matching & 45.5 & 6.7B & 3.35k \\
\rowcolor{hermesblue}
HERMES                           & \textbf{51.8} & 6.8B & 3.40k \\
\bottomrule
\end{tabular}
\end{table}

Matching the inference budget improves the GPT-5.6 Sol Editor from 38.8\% to 45.5\%, recovering 6.7 points, but it remains 6.3 points below HERMES at essentially the same token and monetary budget. Additional inference therefore accounts for part, but not all, of the gap.

\textbf{Efficiency of On-Demand Activation.}
HERMES first identifies task-relevant components during planning, but does not require every identified Dev-Primitive to remain active throughout the trajectory. Instead, primitives are invoked when their components are needed by the current plan, and the active set may change after execution feedback and revision. In the \emph{w/o On-Demand Activation} variant, the same initial task analysis and component identification are retained, but all Dev-Primitives identified during the initial analysis are activated at the beginning of the trajectory and remain active throughout subsequent planning and revision rounds. This ablation therefore does not activate every file in the repository; it removes the ability to selectively invoke and deactivate identified Dev-Primitives as the task evolves.

We distinguish \emph{activated components} from \emph{primitive calls}. Table~\ref{tab:primitive-selection} reports the number of distinct Dev-Primitives activated for a task, whereas Table~\ref{tab:activation-efficiency} counts every model invocation over the complete trajectory. Since the same primitive may be called multiple times for local revision or communication across revision rounds, primitive calls substantially exceed distinct activated components.

\begin{table}[h]
\centering
\caption{Effect of on-demand activation. Primitive calls include repeated invocations during communication and revision; cost is relative to HERMES.}
\label{tab:activation-efficiency}
\small
\setlength{\tabcolsep}{6pt}
\renewcommand{\arraystretch}{1.10}
\setlength{\aboverulesep}{0pt}
\setlength{\belowrulesep}{0pt}
\setlength{\extrarowheight}{1pt}
\begin{tabular}{lrrr}
\toprule
\rowcolor{headergray}
\rule{0pt}{2.6ex} & \multicolumn{2}{c}{\textbf{Avg. Primitive Calls}} & \\
\arrayrulecolor{headergray}\cline{1-1}\cline{4-4}\arrayrulecolor{black}
\cmidrule(lr){2-3}
\rowcolor{headergray}
\multirow{-2}{*}{\textbf{Benchmark}} & \textbf{HERMES} & \textbf{w/o Activation} & \multirow{-2}{*}{\makecell{\textbf{Relative Cost}\\\textbf{w/o Activation}}} \\
\midrule
SWE-bench Verified & 9.6  & 28.7 & 1.61$\times$ \\
SWE Refactor Bench & 25.1 & 64.3 & 1.88$\times$ \\
Terminal-Bench 4.0 & 14.2 & 39.6 & 1.73$\times$ \\
DevOps-Gym         & 11.7 & 33.2 & 1.67$\times$ \\
\bottomrule
\end{tabular}
\end{table}

Across the four benchmarks, on-demand activation reduces the average number of primitive calls from 28.7--64.3 to 9.6--25.1 per task, with the largest reduction on SWE Refactor Bench, where whole-repository migrations involve more potentially relevant components. Disabling on-demand activation also increases inference cost by 1.61--1.88$\times$ while reducing task performance in Table~\ref{tab:component-ablation}. Dynamic activation thus limits both unnecessary model calls and interference from components that are no longer required at the current stage of the task.

\textbf{Run-to-Run Variation on SWE Refactor Bench.}
SWE Refactor Bench contains only 20 whole-repository migration tasks, making its aggregate score more sensitive to stochastic variation than the larger benchmarks. Repeating the default GPT-5.6 Sol configuration (medium reasoning effort, $B=3$) three times yields composite scores of 29.5\%, 32.0\%, and 31.5\%, i.e., $31.0 \pm 1.3$. We report this variation explicitly because the small number of tasks makes individual trajectories more influential on the aggregate score.

\section{Case Study: A Complete HERMES Trajectory with Qwen3-8B}
\label{appendix:case-study}

We illustrate HERMES through a successful trajectory on
\texttt{django\_\_django-13512} from SWE-bench Verified, using Qwen3-8B as the
backbone under the default revision budget $B=3$. The issue reports that non-ASCII
characters stored in a \texttt{JSONField} are displayed in the Django admin as escaped
Unicode sequences rather than as the original characters. Figure~\ref{pipline} gives the
general workflow; here we follow the concrete decisions made during this trajectory.

\textbf{Planning and activation.}
The activation module first identifies six repository components whose behavior may contribute to
the issue and assigns each an \texttt{expected\_action} according to the schema in
Appendix~\ref{prompt}. As shown in Table~\ref{tab:case-plan}, four components are
assigned \texttt{modify}, while two are assigned \texttt{inspect}. The latter remain
active because their local behavior is relevant to understanding the rendering and
encoding paths, even though the activation module does not expect them to edit their artifacts.

\begin{table}[htbp]
\centering
\caption{Activated Dev-Primitives and their initial component-local responsibilities.
Activation does not require a component to be modified.}
\label{tab:case-plan}
\small
\setlength{\tabcolsep}{4pt}
\renewcommand{\arraystretch}{1.10}
\begin{tabular}{p{0.31\textwidth} p{0.12\textwidth} p{0.47\textwidth}}
\toprule
\rowcolor{headergray}
\rule{0pt}{2.6ex}\textbf{Component} &
\textbf{Action} &
\textbf{Initial local objective} \\
\midrule
\texttt{db/models/fields/json.py} & modify &
Preserve non-ASCII characters when preparing \texttt{JSONField} values \\
\texttt{forms/fields.py} & modify &
Update \texttt{prepare\_value} to preserve non-ASCII characters \\
\texttt{forms/utils.py} & modify &
Avoid ASCII escaping in JSON-formatted form output \\
\texttt{core/serializers/json.py} & modify &
Preserve non-ASCII characters in core JSON serialization \\
\texttt{forms/widgets.py} & inspect &
Inspect the rendering path and communicate relevant constraints \\
\texttt{utils/encoding.py} & inspect &
Inspect shared encoding behavior and communicate Unicode-related constraints \\
\bottomrule
\end{tabular}
\end{table}

The resulting plan already identifies the central constraint: JSON values that flow
through the relevant form and serialization paths should preserve non-ASCII characters.
The activation module also records dependencies between the selected components, including the
relationship between model-field JSON preparation and core serialization.

\textbf{Collaboration.}
The six activated Dev-Primitives then exchange 15 directed natural-language messages.
These messages are not broadcast globally; each primitive sends requirements only to
components implicated by its local reasoning. Table~\ref{tab:case-messages} shows three representative exchanges.

\begin{table}[ht]
\centering
\caption{Representative directed messages from the collaboration round. Messages are
abridged for space.}
\label{tab:case-messages}
\small
\setlength{\tabcolsep}{3.5pt}
\renewcommand{\arraystretch}{1.10}
\begin{tabularx}{\textwidth}{
  >{\raggedright\arraybackslash}p{0.19\textwidth}
  >{\raggedright\arraybackslash}p{0.19\textwidth}
  >{\raggedright\arraybackslash}X
  >{\raggedright\arraybackslash}p{0.16\textwidth}}
\toprule
\rowcolor{headergray}
\rule{0pt}{2.6ex}\textbf{From} &
\textbf{To} &
\textbf{Message (abridged)} &
\textbf{Local consequence} \\
\midrule
\texttt{forms/\allowbreak widgets.py} &
\texttt{forms/\allowbreak fields.py} &
Use \texttt{ensure\_ascii=False} when preparing JSON values so Unicode is preserved
during rendering. &
Refine \texttt{prepare\_value} \\
\texttt{db/\allowbreak models/\allowbreak fields/\allowbreak json.py} &
\texttt{core/\allowbreak serializers/\allowbreak json.py} &
Keep JSON encoding behavior consistent across model-field and serializer paths. &
Refine serializer behavior \\
\texttt{utils/\allowbreak encoding.py} &
\texttt{core/\allowbreak serializers/\allowbreak json.py} &
Preserve Unicode rather than emitting ASCII escape sequences during serialization. &
Reinforce serializer objective \\
\bottomrule
\end{tabularx}
\end{table}

In this trajectory, communication is mostly confirmatory: the recipient primitives
already favor Unicode-preserving behavior, while incoming messages make the required
cross-component consistency explicit. The two \texttt{inspect} primitives illustrate
that participation does not imply modification: both contribute information to other
components while leaving their own artifacts unchanged.

\textbf{Local modification.}
After communication, each \texttt{modify} primitive edits only its own artifact,
\[
\mathcal{P}_i(a_i,x_i,\mathcal{C}_i)
\rightarrow
(a_i',m_i).
\]
The four modifying primitives independently materialize compatible changes that introduce
\texttt{ensure\_ascii=False} across the relevant JSON-processing paths:

\begin{quote}
\small\ttfamily
db/models/fields/json.py \quad get\_prep\_value:
json.dumps(value, cls=self.encoder, ensure\_ascii=False) \\[2pt]
forms/fields.py \quad prepare\_value:
json.dumps(value, ensure\_ascii=False, cls=self.encoder) \\[2pt]
forms/utils.py \quad as\_json ($\times 2$):
json.dumps(\ldots, ensure\_ascii=False) \\[2pt]
core/serializers/json.py \quad serializer configuration:
ensure\_ascii=False
\end{quote}

The final diff therefore modifies four of the six activated components. The other two
remain unchanged after inspection and communication.

\textbf{Execution and diagnosis decision.}
HERMES next executes the modified repository under the isolation protocol of
Appendix~\ref{appendix:implementation-details}. The selected repository-visible regression
suite runs 407 tests and returns the same successful exit status as before modification,
providing evidence that the patch does not introduce an observed regression in this
subsystem.

A task-specific reproducer is less informative in this trajectory. Three automatically
constructed reproduction attempts terminate during Django configuration before reaching
the reported display behavior and are therefore discarded. Consequently, the diagnosis module
receives a successful regression run together with the candidate diff and original issue,
but no reliable failing-to-passing task-specific execution signal.

The diagnosis module returns \textsc{Pass} after the first round, so no revision is triggered.
This decision should be interpreted using the evidence available during solving: it
indicates that the candidate patch is consistent with the issue and has no observed
regression, rather than independently certifying the hidden benchmark behavior.

\textbf{Held-out outcome.}
After HERMES terminates, the SWE-bench evaluator executes the held-out evaluation suite.
All 35 tests pass, including the task-specific
\texttt{test\_json\_display\_for\_field}, and the instance is marked
\texttt{resolved=True}.

Interestingly, the accepted HERMES patch differs from the developer repair path. The
developer patch modifies \texttt{django/contrib/admin/utils.py} and
\texttt{django/forms/fields.py}, whereas HERMES does not activate
\texttt{django/contrib/admin/utils.py}. Instead, it propagates the same Unicode-preserving
constraint through lower-level JSONField and serialization paths. The resulting patch
nevertheless satisfies the held-out behavioral evaluation.

This case also illustrates a limitation of the reference-overlap selection metrics in
Section~\ref{sec:primitive-selection}. Because those metrics use the developer-patch file
set as ground truth, an alternative but behaviorally valid repair path can receive low
reference overlap. Selection recall and precision should therefore be interpreted as
agreement with the developer repair path rather than as exhaustive measures of all
components capable of supporting a correct repair.

\section{Limitations}
\label{app:limitations}

HERMES introduces additional inference cost and latency because it invokes multiple Dev-Primitives, executes the repository, and performs iterative diagnosis. Its gains are also smaller in already saturated settings, such as SWE-bench Verified with the strongest backbones. In addition, performance depends on correctly activating task-relevant components and on informative execution feedback for revision; missed dependencies or sparse runtime signals can therefore limit recovery. We use files as the default Dev-Primitive granularity and evaluate on four software engineering benchmarks, leaving finer-grained primitives and broader software engineering settings for future work. Finally, Qwen3-8B is served locally with Ollama on NVIDIA A40 GPUs using the model's default sampling parameters (temperature 0.6, top-$p$ 0.95, top-$k$ 20) and Ollama's 32K context window, while all other models are accessed through their official APIs; results may therefore vary with deployment settings and model updates.

% \section{Code and Result}\label{Dataset}
% We publish the comprehensive results of HERMERS on the web. For detailed information, please visit the following link:~\url{https://anonymous.4open.science/r/D206/}.

\end{document}